\documentclass[sn-mathphys-num]{sn-jnl}

\usepackage{lmodern}
\usepackage{graphicx}%
\usepackage{multirow}%
\usepackage{amsmath,amssymb,amsfonts}%
\usepackage{amsthm}%
\usepackage{mathrsfs}%
\usepackage[title]{appendix}%
\usepackage{xcolor}%
\usepackage{textcomp}%
\usepackage{manyfoot}%
\usepackage{booktabs}%
\usepackage{algorithm}%
\usepackage{algorithmicx}%
\usepackage{algpseudocode}%
\usepackage{listings}%
\usepackage{array}
\usepackage{appendix}
\usepackage{colortbl}
\usepackage{ragged2e}
\usepackage{caption}

\begin{document}

\title[Article Title]{Unified Framework for Bidirectional and Cyclic Teleportation under Noise}


\author[1]{\fnm{Soubhik} \sur{De}}\email{soubhik.de11@pondiuni.ac.in}
\author[1,]{\fnm{Vedhanayagi} \sur{R}}\email{vedhaphy@pondiuni.ac.in}
\author[2,]{\fnm{A Basherrudin Mahmud} \sur{Ahmed}}\email{abmahmed.physics@mkuniversity.org}
\author*[1,]{\fnm{Alok} \sur{Sharan}}\email{aloksharan@pondiuni.ac.in}

\affil[1]{\orgdiv{Department of Physics}, \orgname{Pondicherry University}, \orgaddress{ \city{Puducherry}, \postcode{605014},  \country{India}}}

\affil[2]{\orgdiv{Department of Theoretical Physics, School of Physics}, \orgname{Madurai Kamaraj University}, \orgaddress{ \city{Madurai}, \postcode{625021}, \state{Tamil Nadu}, \country{India}}}

\abstract{
Quantum teleportation has evolved from single-qubit, unidirectional communication to multi-qubit and multidirectional protocols. However, most existing schemes rely on protocol-specific entangled resources, motivating the development of universal quantum channels capable of supporting multiple communication tasks simultaneously. In this work, we demonstrate that a single twelve-qubit entangled channel exhibits such versatility by enabling the bidirectional teleportation of arbitrary three-qubit states and the cyclic teleportation of arbitrary two-qubit states through local Bell-state measurements and single-qubit operations. Both protocols are further generalized to multi-qubit and multi-party configurations, establishing the scalability of the proposed framework. To assess its practical applicability, the protocols are analyzed under amplitude-damping, phase-damping, bit-flip, phase-flip, and depolarizing noise channels by plotting the teleportation fidelity as a function of both the input-state parameters and noise strength. The analysis reveals distinct noise sensitivities, with the bidirectional protocol remaining perfectly faithful under bit-flip noise for all input states and noise strengths, while the cyclic protocol is consistently more vulnerable to environmental disturbances. The proposed schemes achieve an intrinsic efficiency of $25\%$, which is compared with several existing protocols. The framework therefore provides a scalable and resource-efficient approach to unified quantum communication in realistic noisy quantum networks.
}




\maketitle

\section{Introduction}
Quantum teleportation (QT) enables the transfer of unknown quantum states between distant parties through shared entanglement and classical communication. Following its theoretical proposal and experimental realization \cite{bennett_teleporting_1993,bouwmeester_experimental_1997,nielsen_quantum_2010}, bidirectional quantum teleportation (BQT) extended conventional teleportation to simultaneous two-way communication between users, with early protocols demonstrating probabilistic state exchange \cite{mishra_two-way_2011}. Subsequent developments introduced controlled BQT using cluster states \cite{zha_bidirectional_2013}, followed by a general framework for identifying suitable multi-qubit entangled channels that unified a broad class of controlled teleportation schemes \cite{thapliyal_general_2015}. The scope of BQT was further expanded to the teleportation of quantum operations \cite{zhou_bidirectional_2018}, multi-qubit communication through six-qubit entangled channels \cite{zhou_quantum_2020}, and controlled asymmetric protocols that enable flexible allocation of communication roles among participants \cite{huo_controlled_2021}. More recently, resumable teleportation strategies have been proposed to improve the practicality and robustness of bidirectional quantum communication \cite{kumar_mandal_resumable_2024}. Cyclic quantum teleportation (CYQT) further generalized quantum communication to loop-like multipartite networks, where quantum states are transferred sequentially among adjacent users \cite{chen_cyclic_2017}. Subsequent developments expanded cyclic quantum teleportation in several directions. Controller parties were first incorporated to enhance security and coordination \cite{sang_cyclic_2018}, followed by the use of genuinely multipartite-entangled \cite{shao_circular_2019,wang_synchronous_2022}, GHZ-like \cite{verma_cyclic_2020}, and cluster states \cite{verma_cyclic_2021,slaoui_cyclic_2024} as quantum channels. The protocol was further extended to asymmetric communication \cite{yang_quantum_2022,yang_asymmetric_2022}, including controlled asymmetric schemes \cite{choudhury_controlled_2020}, while resource-efficient implementations were subsequently proposed \cite{verma_symmetric_2021,kaur_asymmetric_2024}. More recently, hierarchical controlled cyclic teleportation protocols have also been investigated \cite{zha_hierarchical_2019,zha_different_2025}.

Recent studies have investigated resource-efficient entanglement quantum channels for hybrid communication protocols \cite{zha_four-directional_2020,fatahi_multi-hop_2024,saha_hybrid_2025,paulson_bounds_2017}, demonstrating that distinct quantum communication tasks can be integrated within a unified framework \cite{zhou_cyclic_2019,wu_cyclic_2020,verma_bidirectional_2020}. However, most existing approaches remain protocol-specific, requiring distinct entangled channels for different communication tasks. 
To address this limitation, we propose a unified teleportation framework that implements both bidirectional and cyclic protocols using a single entangled resource. In this framework, the bidirectional configuration enables simultaneous exchange of arbitrary three-qubit states between two parties, while the cyclic configuration realizes transfer of general two-qubit states among three users in a loop.
For this purpose, we employ a twelve-qubit entangled resource state whose structure simultaneously encodes the correlations required for both configurations.
The framework is further generalized to accommodate multiple participants using simple single-qubit gates and Bell-state measurements, enabling exchange of multi-qubit states across larger quantum networks with multiple participants.

Practical quantum systems are inherently subject to noise due to environmental interactions, leading to decoherence and degradation of quantum information. The impact of noise on teleportation protocols has been extensively studied for both bidirectional and cyclic schemes \cite{xu_asymmetric_2023,choudhury_bi-directional_2023,liu_cyclic-controlled_2019,kaur_multidirectional_2023,sisodia_hybrid_2024}. Various mitigation strategies, including weak measurement techniques and optimized channel design, have been proposed to preserve teleportation fidelity \cite{yang_bidirectional_2017,mandal_quantum_2024,kaur_efficient_2024, taufiqi_teleportation_2025}.
In this work, we investigate the effects of amplitude-damping, phase-damping, bit-flip, phase-flip, and depolarizing noises acting on the quantum channel by analyzing the teleportation fidelity of both the bidirectional and cyclic protocols as functions of the input-state parameters and noise strength $p$.
Heat-map visualizations identify regimes of high-fidelity operation, while comparative analysis under identical noise conditions reveals their relative robustness and the classes of states least affected by specific noise models. Finally, the intrinsic efficiency of the proposed schemes is evaluated and compared with existing protocols to demonstrate the advantages of the framework in terms of resource utilization.

The remainder of the paper is organized as follows: Section \ref{sec_4} details the construction of the quantum channel to be used for the unified framework. Sections \ref{sec_1} and \ref{sec_2} describe the bidirectional and cyclic teleportation protocols, respectively. Section \ref{sec_5} generalizes the framework to variable number of qubits and arbitrary participants, starting with the generalization of the quantum channel. Section \ref{sec_6} analyzes the effects of different noise models on teleportation fidelity, while Section \ref{sec_7} evaluates the intrinsic efficiency of the proposed schemes and compares them with existing protocols. Section \ref{sec_8} gives the concluding remarks and highlights the future directions of the work.

\section{Construction of the quantum channel}
\label{sec_4} The detailed steps of our twelve-qubit quantum channel is given as follows: \\[5pt]
\textit{Step 1}. Prepare twelve qubits in the state $|0\rangle$ as initial state:
\begin{equation}
    |0\rangle_1 \otimes |0\rangle_2 \otimes|0\rangle_3 \otimes|0\rangle_4 \otimes|0\rangle_5 \otimes|0\rangle_6 \otimes |0\rangle_7\otimes|0\rangle_8 \otimes|0\rangle_9 \otimes|0\rangle_{10} \otimes|0\rangle_{11} \otimes|0\rangle_{12} 
\end{equation}
\textit{Step 2}. Apply six Hadamard gates to the qubits $1,2,3,4,5,6$ respectively, which leads to the overall quantum state:
\begin{align}
        \frac{1}{\sqrt{2}}(|0\rangle +|1\rangle)_1 \otimes \frac{1}{\sqrt{2}}(|0\rangle +|1\rangle)_2 \otimes \frac{1}{\sqrt{2}}(|0\rangle +|1\rangle)_3 \otimes \frac{1}{\sqrt{2}}(|0\rangle +|1\rangle)_4 \otimes \nonumber \\\frac{1}{\sqrt{2}}(|0\rangle +|1\rangle)_5 \otimes \frac{1}{\sqrt{2}}(|0\rangle +|1\rangle)_6 \otimes |0\rangle_7 \otimes |0\rangle_8 \otimes |0\rangle_9 \otimes |0\rangle_{10} \otimes |0\rangle_{11} \otimes |0\rangle_{12}
\end{align}
\textit{Step 3}. Implement six CNOT gate operations on the qubit pairs $(1,7)$, $(2,8)$, $(3,9)$, $(4,10)$, $(5,11)$, and $(6,12)$ respectively, where the first qubit acts as control and second qubit as target. This will generate the twelve-qubit resource channel state:
\begin{align}
    \label{eqn_1}
    |G_{12}\rangle = \frac{1}{8}
    \big(
        |000000000000\rangle +
        |000001000001\rangle + 
        |000010000010\rangle + 
        |000011000011\rangle + \nonumber \\[-4.5pt]
        |000100000100\rangle + 
        |000101000101\rangle + 
        |000110000110\rangle + 
        |000111000111\rangle + \nonumber \\
        |001000001000\rangle + 
        |001001001001\rangle + 
        |001010001010\rangle + 
        |001011001011\rangle + \nonumber \\
        |001100001100\rangle + 
        |001101001101\rangle + 
        |001110001110\rangle +
        |001111001111\rangle +\nonumber \\
        |010000010000\rangle + 
        |010001010001\rangle + 
        |010010010010\rangle + 
        |010011010011\rangle +\nonumber \\
        |010100010100\rangle + 
        |010101010101\rangle + 
        |010110010110\rangle + 
        |010111010111\rangle +\nonumber \\
        |011000011000\rangle +
        |011001011001\rangle + 
        |011010011010\rangle + 
        |011011011011\rangle + \nonumber \\
        |011100011100\rangle + 
        |011101011101\rangle +
        |011110011110\rangle + 
        |011111011111\rangle +\nonumber \\
        |100000100000\rangle + 
        |100001100001\rangle + 
        |100010100010\rangle +
        |100011100011\rangle + \nonumber \\
        |100100100100\rangle + 
        |100101100101\rangle + 
        |100110100110\rangle + 
        |100111100111\rangle + \nonumber \\
        |101000101000\rangle + 
        |101001101001\rangle + 
        |101010101010\rangle + 
        |101011101011\rangle + \nonumber \\
        |101100101100\rangle +
        |101101101101\rangle + 
        |101110101110\rangle + 
        |101111101111\rangle +\nonumber \\
        |110000110000\rangle + 
        |110001110001\rangle +
        |110010110010\rangle + 
        |110011110011\rangle + \nonumber \\
        |110100110100\rangle + 
        |110101110101\rangle + 
        |110110110110\rangle +
        |110111110111\rangle +\nonumber \\
        |111000111000\rangle + 
        |111001111001\rangle + 
        |111010111010\rangle + 
        |111011111011\rangle +\nonumber \\
        |111100111100\rangle + 
        |111101111101\rangle + 
        |111110111110\rangle + 
        |111111111111\rangle
    \big)
\end{align}

\section{Bidirectional teleportation $3 \Leftrightarrow 3$}
\label{sec_1}
Consider two parties, Alice and Bob, who wish to exchange unknown quantum states with one another. The unknown three-qubit message state with Alice can be expressed as:
\begin{align}
    \label{eqn_3}
    |\Psi_A\rangle_{a_1a_2a_3} = \alpha_0|000\rangle + \alpha_1 |001\rangle + \alpha_2|010\rangle +\alpha_3|011\rangle + \alpha_4|100\rangle + \alpha_5 |101\rangle \nonumber \\ 
    + \alpha_6 |110\rangle + \alpha_7|111\rangle
\end{align}
Similarly, Bob's message state is given by:
\begin{align}
    |\Psi_B\rangle_{b_1b_2b_3} = \beta_0|000\rangle + \beta_1 |001\rangle + \beta_2|010\rangle +\beta_3|011\rangle + \beta_4|100\rangle + \beta_5 |101\rangle \nonumber \\ 
    + \beta_6 |110\rangle + \beta_7 |111\rangle 
\end{align}
where complex coefficients $\sum_i |\alpha_i|^2=1$ and $\sum_j |\beta_j|^2=1$.

The channel state in Eqn. \ref{eqn_1} is distributed between Alice and Bob such that the qubits $1,2,3,10,11,12$ belong to Alice and qubits $4,5,6,7,8,9$ belong to Bob, ensuring that the participants share maximal entanglement between them. The overall state of the system is thus given by:
\begin{equation}
    |\Phi^{(3)}\rangle = |\Psi_A\rangle_{a_1a_2a_3} \otimes |\Psi_B\rangle_{b_1b_2b_3} \otimes |G_{12}\rangle_{1-12}
\end{equation}
After distribution of the qubits to corresponding parties, the protocol is carried out:

\vspace{5pt}
{\justifying
\textit{Step 1}. Alice performs a Bell-state measurement (BSM) on the qubit pairs $(a_1,1)$. The measurement basis of the BSM is given by:
\begin{equation}
    |\phi^\pm\rangle=\frac{1}{\sqrt{2}}(|00\rangle\pm|11\rangle), \quad |\psi^\pm\rangle=\frac{1}{\sqrt{2}}(|01\rangle\pm|10\rangle)
\end{equation}
In terms of this Bell basis, the state $|\Phi^{(3)}\rangle$ can be rewritten as:
\begin{align}
    \frac{1}{\sqrt2} \Big\{|\phi^+\rangle_{a_1,1}\otimes\big(|\lambda_0\rangle_{a_2,a_3}|0\rangle_{7} + |\lambda_4\rangle_{a_2,a_3}|1\rangle_{7}\big) + |\phi^-\rangle_{a_1,1}\otimes \big(|\lambda_0\rangle_{a_2,a_3}|0\rangle_{7} \nonumber \\ 
    -|\lambda_4\rangle_{a_2,a_3}|1\rangle_{7}\big) + |\psi^+\rangle_{a_1,1}\otimes \big(|\lambda_0\rangle_{a_2,a_3}|1\rangle_{7} + |\lambda_4\rangle_{a_2,a_3}|0\rangle_{7}\big) + |\psi^-\rangle_{a_1,1} \nonumber \\
    \otimes\big(|\lambda_0\rangle_{a_2,a_3}|1\rangle_{7} - |\lambda_4\rangle_{a_2,a_3}|0\rangle_{7}\big)\Big\} \otimes |\Psi^{(3)}_B\rangle_{b_1b_2b_3} \otimes |G_{10}\rangle_{2-6,~8-12}
\end{align}
where
\begin{align}
    |\lambda_0\rangle = \alpha_0|00\rangle+\alpha_1|01\rangle+\alpha_2|10\rangle+\alpha_3|11\rangle, \nonumber \\
    |\lambda_4\rangle = \alpha_4|00\rangle+\alpha_5|01\rangle+\alpha_6|10\rangle+\alpha_7|11\rangle
\end{align}
and
\begin{align}
    \label{eqn_9}
    |G_{10}\rangle = \frac{1}{4\sqrt2}\big( 
        |0000000000\rangle + |0000100001\rangle + |0001000010\rangle + |0001100011\rangle + \nonumber \\[-5.5pt]
        |0010000100\rangle + |0010100101\rangle + |0011000110\rangle + |0011100111\rangle + \nonumber \\
        |0100001000\rangle + |0100101001\rangle + |0101001010\rangle + |0101101011\rangle + \nonumber \\
        |0110001100\rangle + |0110101101\rangle + |0111001110\rangle + |0111101111\rangle + \nonumber \\
        |1000010000\rangle + |1000110001\rangle + |1001010010\rangle + |1001110011\rangle + \nonumber \\
        |1010010100\rangle + |1010110101\rangle + |1011010110\rangle + |1011110111\rangle + \nonumber \\
        |1100011000\rangle + |1100111001\rangle + |1101011010\rangle + |1101111011\rangle + \nonumber \\
        |1110011100\rangle + |1110111101\rangle + |1111011110\rangle + |1111111111\rangle 
         \big)
\end{align}
If Alice's first BSM result is $|\psi^+\rangle$, they are left with the collapsed state:
\begin{align}
    |\Phi^{\prime(3)}\rangle = \frac{1}{8} \Big\{ |\psi^+\rangle_{a_1,1}\otimes\big(|\lambda_0\rangle_{a_2,a_3}|1\rangle_{7} + |\lambda_4\rangle_{a_2,a_3}|0\rangle_{7}\big) \otimes |\Psi^{(3)}_B\rangle_{b_1b_2b_3} \otimes |G_{10}\rangle_{2-6,8-12}\Big\}
\end{align}
Alice performs her second BSM on qubits $(a_2,2)$. The state $|\Phi^{\prime(3)}\rangle$ can be rewritten in terms of Bell basis as:
\begin{align}
    \frac{1}{8} |\psi^+\rangle_{a_1,1} \otimes\Big\{|\phi^+\rangle_{a_2,2}\otimes\big(|\mu_0\rangle_{a_3}|10\rangle_{7,8} + |\mu_4\rangle_{a_3}|00\rangle_{7,8} + |\mu_2\rangle_{a_3}|11\rangle_{7,8} \nonumber \\
    +|\mu_6\rangle_{a_3}|01\rangle_{7,8}\big) + |\phi^-\rangle_{a_2,2}\otimes\big(|\mu_0\rangle_{a_3}|10\rangle_{7,8} + |\mu_4\rangle_{a_3}|00\rangle_{7,8} \nonumber \\
    - |\mu_2\rangle_{a_3}|11\rangle_{7,8} - |\mu_6\rangle_{a_3}|01\rangle_{7,8}\big)  + |\psi^+\rangle_{a_2,2}\otimes \big(|\mu_0\rangle_{a_3}|11\rangle_{7,8} \nonumber \\
    + |\mu_4\rangle_{a_3}|01\rangle_{7,8} + |\mu_2\rangle_{a_3}|10\rangle_{7,8} + |\mu_6\rangle_{a_3}|00\rangle_{7,8}\big) + |\psi^-\rangle_{a_2,2} \nonumber \\
    \otimes \big(|\mu_0\rangle_{a_3}|11\rangle_{7,8} + |\mu_4\rangle_{a_3}|01\rangle_{7,8} - |\mu_2\rangle_{a_3}|10\rangle_{7,8} \nonumber \\
    - |\mu_6\rangle_{a_3}|00\rangle_{7,8}\big)\Big\} \otimes |\Psi^{(3)}_B\rangle_{b_1b_2b_3} \otimes |G_{8}\rangle_{3-6,9-12}
\end{align}
where
\begin{align}
    \label{eqn_8}
    &|\mu_0\rangle = \alpha_0|0\rangle + \alpha_1|1\rangle, \quad |\mu_2\rangle = \alpha_2|0\rangle + \alpha_3|1\rangle, \nonumber \\
    &|\mu_4\rangle = \alpha_4|0\rangle + \alpha_5|1\rangle, \quad |\mu_6\rangle = \alpha_6|0\rangle + \alpha_7|1\rangle
\end{align}
and
\begin{align}
    \label{eqn_10}
    |G_{8}\rangle = \frac{1}{4}\big(
     |00000000\rangle
    +|00010001\rangle 
    +|00100010\rangle 
    +|00110011\rangle 
    +|01000100\rangle \nonumber \\[-4.5pt]
    +|01010101\rangle 
    +|01100110\rangle 
    +|01110111\rangle
    +|10001000\rangle 
    +|10011001\rangle \nonumber \\
    +|10101010\rangle 
    +|10111011\rangle
    +|11001100\rangle 
    +|11011101\rangle 
    +|11101110\rangle \nonumber \\
    +|11111111\rangle
    \big)
\end{align}
If Alice's second BSM result is $|\phi^-\rangle$, then the collapsed state is:
\begin{align}
    |\Phi^{\prime\prime(3)}\rangle=\frac{1}{8} \Big\{|\psi^+\rangle_{a_1,1} |\phi^-\rangle_{a_2,2} \otimes \big(|\mu_0\rangle_{a_3}|10\rangle_{7,8} -|\mu_2\rangle_{a_3}|11\rangle_{7,8} + |\mu_4\rangle_{a_3}|00\rangle_{7,8} - \nonumber \\ |\mu_6\rangle_{a_3}|01\rangle_{7,8} \big)
    \otimes |\Psi^{(3)}_B\rangle_{b_1b_2b_3} \otimes |G_{8}\rangle_{3-6,9-12}\Big\}
\end{align}
Alice performs her third BSM on qubits $(a_3,3)$, which leads to the state $|\Phi^{\prime\prime(3)}\rangle$ being rewritten in terms of the Bell basis states as:
\begin{align}
    \frac{1}{8} |\psi^+\rangle_{a_1,1} |\phi^-\rangle_{a_2,2} \otimes \Big\{|\phi^+\rangle_{a_3,3} \otimes \big(\alpha_0|100\rangle + \alpha_1|101\rangle - \alpha_2|110\rangle - \alpha_3|111\rangle + \alpha_4|000\rangle \nonumber \\
    + \alpha_5|001\rangle -\alpha_6|010\rangle - \alpha_7|011\rangle\big)_{7,8,9} + |\phi^-\rangle_{a_3,3} \otimes \big(\alpha_0|100\rangle - \alpha_1|101\rangle - \alpha_2|110\rangle \nonumber \\
    + \alpha_3|111\rangle + \alpha_4|000\rangle -\alpha_5|001\rangle - \alpha_6|010\rangle + \alpha_7|011\rangle\big)_{7,8,9} + |\psi^+\rangle_{a_3,3} \otimes \big(\alpha_0|101\rangle \nonumber \\
    + \alpha_1|100\rangle - \alpha_2|111\rangle - \alpha_3|110\rangle + \alpha_4|001\rangle + \alpha_5|000\rangle - \alpha_6|011\rangle - \alpha_7|010\rangle \big)_{7,8,9} \nonumber \\
    + |\psi^-\rangle_{a_3,3} \otimes \big(\alpha_0|101\rangle - \alpha_1|100\rangle - \alpha_2|111\rangle + \alpha_3|110\rangle + \alpha_4|001\rangle - \alpha_5|000\rangle \nonumber \\
    - \alpha_6|011\rangle + \alpha_7|010\rangle\big)_{7,8,9} \Big\} \otimes |\Psi^{(3)}_B\rangle_{b_1b_2b_3} \otimes |G_{6}\rangle_{4-6,10-12}
\end{align}
where
\begin{align}
    |G_{6}\rangle = \frac{1}{2\sqrt2}\big(&|000000\rangle+|001001\rangle +|010010\rangle +|011011\rangle \nonumber \\
    +&|100100\rangle +|101101\rangle +|110110\rangle +|111111\rangle \big)
\end{align}
If Alice measures the state to be $|\psi^-\rangle$, then the collapsed state is:
\begin{align}
    |\Phi^{\prime\prime\prime(3)}\rangle = \frac{1}{8} \Big\{|\psi^+\rangle_{a_1,1} |\phi^-\rangle_{a_2,2} |\psi^-\rangle_{a_3,3} \otimes \big( \alpha_0|101\rangle - \alpha_1|100\rangle - \alpha_2|111\rangle \nonumber \\
    + \alpha_3|110\rangle + \alpha_4|001\rangle - \alpha_5|000\rangle - \alpha_6|011\rangle + \alpha_7|010\rangle\big)_{7,8,9} \nonumber \\
    \otimes |\Psi^{(3)}_B\rangle_{b_1b_2b_3} \otimes |G_{6}\rangle_{4-6,10-12}\Big\}
\end{align}
Alice shares her measurement results to Bob via classical communication using six classical bits.

\textit{Step 2}. Similarly, Bob performs Bell-state measurements on the qubit pairs $(b_1,4)$, $(b_2,5)$, and $(b_3,6)$. If his measurement results are $|\phi^+\rangle_{b_1,4}$, $|\psi^+\rangle_{b_2,5}$ and $|\phi^-\rangle_{b_3,6}$, then the total collapsed state is given by:
\begin{align}
    |\Phi^{\prime\prime\prime\prime(3)}\rangle = \frac{1}{8} \Big\{|\psi^+\rangle_{a_1,1} |\phi^-\rangle_{a_2,2} |\psi^-\rangle_{a_3,3} \otimes \big(\alpha_0|101\rangle - \alpha_1|100\rangle - \alpha_2|111\rangle \nonumber \\
    + \alpha_3|110\rangle + \alpha_4|001\rangle - \alpha_5|000\rangle - \alpha_6|011\rangle + \alpha_7|010\rangle \big)_{7,8,9} \nonumber \\
    \otimes |\phi^+\rangle_{b_1,4} |\psi^+\rangle_{b_2,5} |\phi^-\rangle_{b_3,6} \otimes \big(\beta_0|010\rangle - \beta_1|011\rangle + \beta_2|000\rangle \nonumber\\
    - \beta_3|001\rangle + \beta_4|110\rangle - \beta_5|111\rangle + \beta_6|100\rangle - \beta_7|101\rangle\big)_{10,11,12}\Big\}
\end{align}
Bob sends his measurement results to Alice using six classical bits via classical channel.

\textit{Step 3}. Alice's measurement results being $|\psi^+\rangle_{a_1,1}$, $|\phi^-\rangle_{a_2,2}$ and $|\psi^-\rangle_{a_3,3}$, Bob applies the corrective operators $X_7$, $Z_8$ and $Z_9X_9$ to get Alice's message state successfully.
Similarly, since Bob's measurement results are $|\phi^+\rangle_{b_1,4}$, $|\psi^+\rangle_{b_2,5}$ and $|\phi^-\rangle_{b_3,6}$, Alice applies the unitaries $I_{10}$, $X_{11}$ and $Z_{12}$ to recover Bob's message state faithfully onto her qubits.
}
\vspace{5pt}

\begin{figure}
    \centering
    \includegraphics[width=\linewidth]{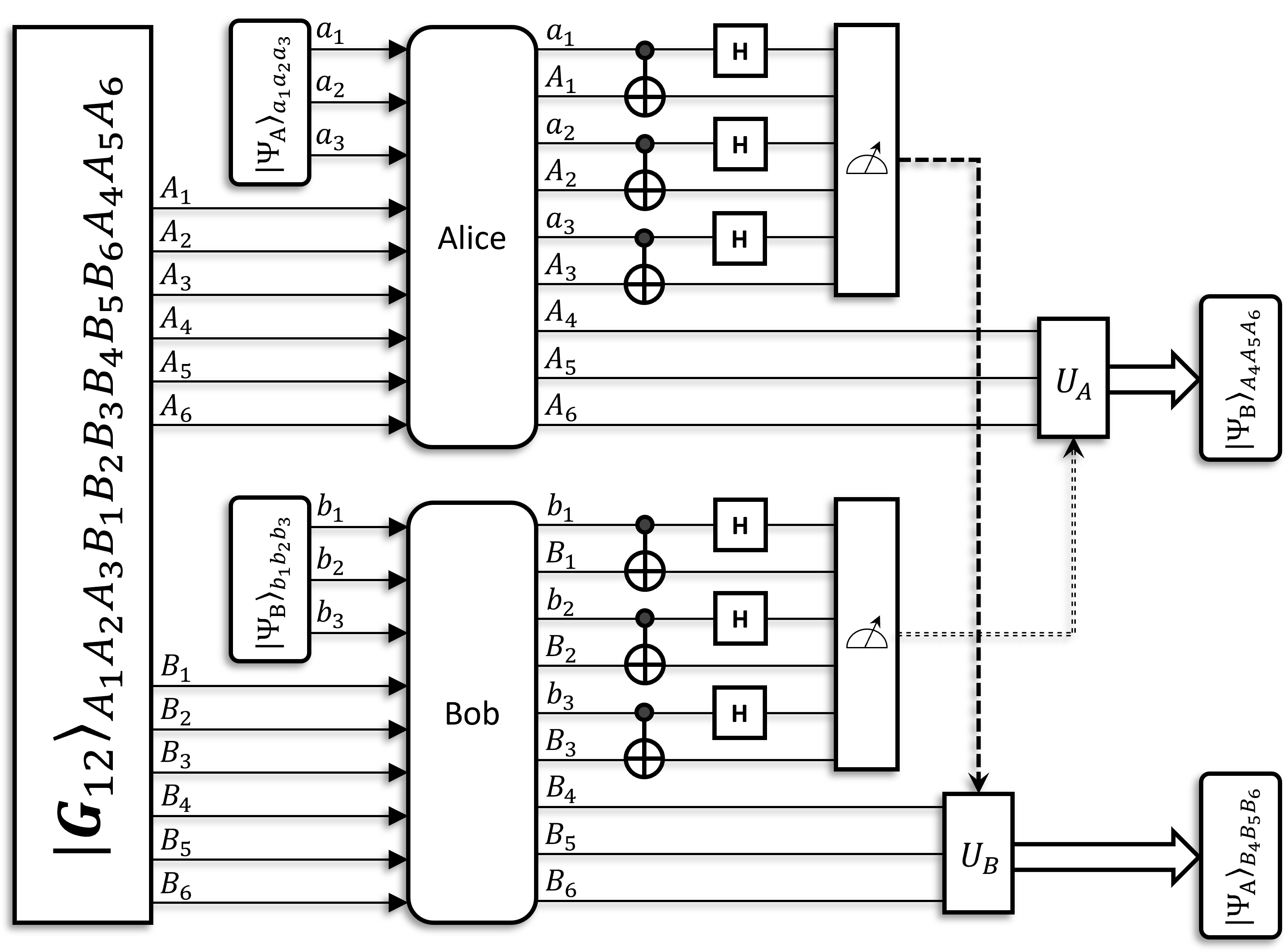}
    \caption{Schematic diagram depicting the bidirectional teleportation of arbitrary three-qubit states via a twelve-qubit entangled state. $U_A$ and $U_B$ represent the unitary operations to be done by Alice and Bob respectively, and dashed arrows represent classical communication.}
    \label{fig_scheme_bqt}
\end{figure}

The protocol gives rise to $2^{12}$ distinct and equiprobable measurement combinations, each associated with a specific set of corrective unitary operations to be performed by the respective participants, ensuring the total success probability of the protocol is unity. The summary of all such possible measurement combinations for each participant, and the corresponding unitaries they have to apply to reconstruct the transmitted state to their qubits is given in Table \ref{tab_3}. 
The schematic for the implementation of the bidirectional protocol of arbitrary three-qubit states is given in Fig. \ref{fig_scheme_bqt}.

\section{\label{sec_2}Cyclic teleportation $2\Leftrightarrow2\Leftrightarrow2$}
Alice, Bob and Charlie are three users, each of whom possess unknown two-qubit states, given by:
\begin{align}
    \label{eqn_7}
    |\Psi_A\rangle_{a_1a_2} &= \alpha_0|00\rangle + \alpha_1|01\rangle + \alpha_2|10\rangle + \alpha_3|11\rangle \nonumber \\[3pt]
    |\Psi_B\rangle_{b_1b_2} &= \beta_0|00\rangle + \beta_1|01\rangle + \beta_2|10\rangle + \beta_3|11\rangle \nonumber \\[3pt]
    |\Psi_C\rangle_{c_1c_2} &= \gamma_0|00\rangle + \gamma_1|01\rangle + \gamma_2|10\rangle + \gamma_3|11\rangle
\end{align}
where $\sum_i |\alpha_i|^2=1$, $\sum_j |\beta_j|^2=1$ and $\sum_k |\gamma_k|^2=1$.

The twelve-qubit channel in Eqn. \ref{eqn_1} is now distributed between Alice, Bob and Charlie in a configuration such that qubits $1,2,11,12$ are with Alice, $3,4,7,8$ are with Bob and $5,6,9,10$ are with Charlie, ensuring maximal entanglement between the parties. The overall state of the system is hence given as:
\begin{equation}
    |\Phi^{(2)}\rangle = |\Psi_A\rangle_{a_1a_2} \otimes |\Psi_B\rangle_{b_1b_2} \otimes |\Psi_C\rangle_{c_1c_2} \otimes |G_{12}\rangle_{1-12}
\end{equation}

\vspace{5pt}
{\justifying
\textit{Step 1}. Alice now performs Bell-state measurement (BSM) on qubits $(a_1,1)$. The state $|\Phi^{(2)}\rangle$ can be rewritten in terms of Bell basis $\{|\phi^\pm\rangle, |\psi^\pm\rangle\}$ as:
\begin{align}
    \frac{1}{8}\Big\{|\phi^+\rangle_{a_1,1}\big(|\mu_0\rangle_{a_2}|0\rangle_7 + |\mu_2\rangle_{a_2}|1\rangle_7\big) + |\phi^-\rangle_{a_1,1}\big(|\mu_0\rangle_{a_2}|0\rangle_7 - |\mu_2\rangle_{a_2}|1\rangle_7\big) + \nonumber \\
    |\psi^+\rangle_{a_1,1}\big(|\mu_0\rangle_{a_2}|1\rangle_7 + |\mu_2\rangle_{a_2}|0\rangle_7\big) + |\psi^-\rangle_{a_1,1}\big(|\mu_0\rangle_{a_2}|1\rangle_7 - |\mu_2\rangle_{a_2}|0\rangle_7\big)\Big\} \nonumber \\
    \otimes |\Psi^{(2)}_B\rangle_{b_1b_2} \otimes |\Psi^{(2)}_C\rangle_{c_1c_2} \otimes |G_{10}\rangle_{2-6,8-12}
\end{align}
where $|\mu_0\rangle$, $|\mu_2\rangle$ are defined in Eqn. \ref{eqn_8}, and $|G_{10}\rangle$ in Eqn. \ref{eqn_9}. Assuming Alice's first BSM result to be $|\phi^-\rangle$, the collapsed state will be given by:
\begin{align}
    |\Phi^{\prime(2)}\rangle = \frac{1}{8} \Big\{ |\phi^-\rangle_{a_1,1}&\otimes\big(|\mu_0\rangle_{a_2}|0\rangle_7 - |\mu_2\rangle_{a_2}|1\rangle_7\big) \nonumber \\
    &\otimes |\Psi^{(2)}_B\rangle_{b_1b_2} \otimes |\Psi^{(2)}_C\rangle_{c_1c_2} \otimes |G_{10}\rangle_{2-6,8-12} \Big\}
\end{align}
Alice performs her second BSM on qubits $(a_2,2)$. The state $|\Phi^{\prime(2)}\rangle$ will be rewritten in terms of Bell states as:
\begin{align}
    \frac{1}{8}|\phi^-\rangle_{a_1,1}\otimes\Big\{|\phi^+\rangle_{a_2,2}\big(|\nu_0\rangle_7|0\rangle_8 + |\nu_1\rangle_7|1\rangle_8\big) + |\phi^-\rangle_{a_2,2}\big(|\nu_0\rangle_7|0\rangle_8 - |\nu_1\rangle_7|1\rangle_8\big) \nonumber \\
    + |\psi^+\rangle_{a_2,2}\big(|\nu_0\rangle_7|1\rangle_8 + |\nu_1\rangle_7|0\rangle_8\big) + |\psi^-\rangle_{a_2,2}\big(|\nu_0\rangle_7|1\rangle_8 + |\nu_1\rangle_7|0\rangle_8\big) \Big\} \nonumber \\
    \otimes |\Psi^{(2)}_B\rangle_{b_1b_2} \otimes |\Psi^{(2)}_C\rangle_{c_1c_2} \otimes|G_8\rangle_{3-6,9-12}
\end{align}
where
\begin{align}
    |\nu_0\rangle = \alpha_0|0\rangle - \alpha_2|1\rangle, \quad
    |\nu_1\rangle = \alpha_1|0\rangle - \alpha_3|1\rangle
\end{align}
and $|G_8\rangle$ is defined in Eqn. \ref{eqn_10}. If Alice measures $|\psi^+\rangle$, then the collapsed state is given by:
\begin{align}
    |\Phi^{\prime\prime(2)}\rangle = \frac{1}{8} \Big\{ |\phi^-\rangle_{a_1,1}|\psi^+\rangle_{a_2,2} \otimes \big(\alpha_0|01\rangle_{7,8}+\alpha_1|00\rangle_{7,8}-\alpha_2|11\rangle_{7,8} \nonumber \\
    -\alpha_3|10\rangle_{7,8}\big) \otimes |\Psi^{(2)}_B\rangle_{b_1b_2} \otimes |\Psi^{(2)}_C\rangle_{c_1c_2} \otimes |G_{8}\rangle_{3-6,9-12} \Big\}
\end{align}
Alice shares her measurement results with her adjacent party Bob, via classical communication using four classical bits. 

\textit{Step 2}.
Similarly, Bob measures $(b_1,3)$ and $(b_2,4)$. If his measurement results are $|\psi^-\rangle_{b_1,3}$ and $|\phi^-\rangle_{b_2,4}$, then the collapsed state will be given by:
\begin{align}
    |\Phi^{\prime\prime\prime(2)}\rangle = \frac{1}{8} \Big\{ |\phi^-\rangle_{a_1,1}|\psi^+\rangle_{a_2,2}|\psi^-\rangle_{b_1,3}|\phi^+\rangle_{b_2,4} \otimes \big(\alpha_0|01\rangle_{7,8}+\alpha_1|00\rangle_{7,8}\nonumber \\
    - ~\alpha_2|11\rangle_{7,8}-\alpha_3|10\rangle_{7,8} \big) \otimes \big( \beta_0 |10\rangle_{9,10} + \beta_1|11\rangle_{9,10} ~- \nonumber \\
     \beta_2|00\rangle_{9,10}-\beta_3 |01\rangle_{9,10} \big) \otimes |\Psi^{(2)}_C\rangle_{c_1c_2} \otimes |G_{4}\rangle_{5-6,11-12} \Big\}
\end{align}
Bob shares his result in four classical bits to the adjacent participant Charlie.

\textit{Step 3}. Charlie performs measurements on $(c_1,5)$ and $(c_2,6)$. The results being $|\phi^+\rangle_{c_1,5}$ and $|\psi^-\rangle_{c_1,6}$, the state collapses to as:
\begin{align}
    |\Phi^{\prime\prime\prime\prime(2)}\rangle = \frac{1}{8} \Big\{ |\phi^-\rangle_{a_1,1}|\psi^+\rangle_{a_2,2}|\psi^-\rangle_{b_1,3}|\phi^+\rangle_{b_2,4}|\phi^+\rangle_{c_1,5}|\psi^-\rangle_{c_2,6} \otimes \big( \alpha_0|01\rangle_{7,8} \nonumber \\
    + \alpha_1|00\rangle_{7,8} - \alpha_2|11\rangle_{7,8} - \alpha_3|10\rangle_{7,8} \big) \otimes \big(\beta_0|10\rangle_{9,10} + \beta_1|11\rangle_{9,10} \nonumber \\
    - \beta_2|00\rangle_{9,10} - \beta_3|01\rangle_{9,10}\big) \otimes \big( \gamma_0|01\rangle_{11,12} - \gamma_1|00\rangle_{11,12} \nonumber \\
    + \gamma_2|11\rangle_{11,12} -\gamma_3|10\rangle_{11,12}\big) \Big\}
\end{align}
Charlie stores his result in four classical bits and sends them to Alice.

\textit{Step 4}. Alice's measurement results are $|\phi^-\rangle_{a_1,1}$ and $|\psi^+\rangle_{a_2,2}$, hence Bob applies the corrective operators $Z_7$ and $X_8$ respectively. Similarly, Bob's results are $|\psi^-\rangle_{b_1,3}$ and $|\phi^+\rangle_{b_2,4}$, leading to Charlie's corrective unitaries $Z_9X_9$ and $I_{10}$, and Charlie's measurement results $|\phi^+\rangle_{c_1,5}$ and $|\psi^-\rangle_{c_2,6}$ give way to Alice applying $I_{11}$ and $Z_{12}X_{12}$ respectively.
}
\vspace{5pt}

These set of BSM outcomes of all the participants and corresponding corrective unitary rotations performed by them to reconstruct the message states produce $2^{12}$ distinct combinations, all of which lead to successful cyclic teleportation in the forward direction (Alice $\Rightarrow$ Bob $\Rightarrow$ Charlie $\Rightarrow$ Alice). The summary of the BSM outcomes along with their corrective unitaries for each participant is mentioned in Table \ref{tab_4}.

We would like to emphasize here that the direction of the proposed cyclic teleportation protocol can be reversed (Alice $\Rightarrow$ Charlie $\Rightarrow$ Bob $\Rightarrow$ Alice) without the need to change the channel state used in the protocol simply by reassignment of the qubits to the participants during the initial distribution. To reverse the direction of cyclic teleportation, Alice takes qubits $1,2,9,10$, Bob has qubits $3,4,11,12$ and Charlie obtains qubits $5,6,7,8$.
The reverse-direction protocol then proceeds analogously, with a change in the measurement outcomes being communicated from Alice $\to$ Charlie, Charlie $\to$ Bob, and Bob $\to$ Alice.

\section{\label{sec_5}Protocol Generalizations}
The bidirectional protocol of Sec. \ref{sec_1} is generalized to arbitrary $n\ge3$-qubit transmission between two parties, while the cyclic protocol of Sec. \ref{sec_2} is extended to $m\ge3$ participants, each transmitting two qubits to a neighbouring node. These generalizations represent two distinct scaling approaches: increasing the quantum information capacity per user in the bidirectional scheme and expanding the network size in the cyclic scheme.

\subsection{Quantum channel generalization}
Both the generalized bidirectional and cyclic protocols are based on the common quantum channel state:
\begin{align}
    \label{eqn_6}
    |G_{4n}\rangle = \frac{1}{2^n} \sum_{\mathbf{x} \in \{0,1\}^{2n}} |\mathbf{x}\mathbf{x}\rangle
\end{align}
where $\mathbf{x} \in \{0,1\}^{2n}$ denotes a binary string of length $2n$, $n$ denoting either the number of transmitted qubits per party in the bidirectional scheme or the number of participants in the cyclic configuration.

\subsection{Bidirectional teleportation $n \Leftrightarrow n$}
The bidirectional protocol is generalized to enable mutual exchange of arbitrary $n$-qubit states between Alice and Bob using the generalized quantum channel. The unknown message states possessed by the two parties are of the form:
\begin{equation}
    |\Psi^n\rangle = \sum_{i = 0}^{2^n-1} \alpha_i\hspace{1pt}|i\rangle,
\end{equation}
where $\sum_i|\alpha_i|^2=1$. The $4n$ qubits of the channel state in Eqn. \ref{eqn_6} are divided into four $n$-qubit subsets, with the first and fourth assigned to Alice and the remaining two assigned to Bob. The protocol then proceeds as follows:

\vspace{5pt}
{\justifying
\textit{Step 1}. Alice and Bob each perform $n$ Bell-state measurements on pairs of message and resource qubits, and exchange the outcomes classically.

\textit{Step 2}. Based on the received $2n$ classical bits, each party applies appropriate Pauli-X and Pauli-Z corrections on their respective qubits.
}
\vspace{5pt}

\subsection{Cyclic teleportation $(2 \Leftrightarrow2)^{(n-1)}$}
The cyclic protocol is generalized to an n-participant network in which each user transfers an arbitrary two-qubit state to an adjacent participant using the generalized quantum channel. The message state of the $I^{th}$ participant is:
\begin{equation}
    |\Psi^2_I\rangle = \alpha^I_0 |00\rangle + \alpha^I_1 |01\rangle + \alpha^I_2 |10\rangle + \alpha^I_3 |11\rangle
\end{equation}
where $\sum_j|\alpha^I_j|^2=1$. The $4n$ qubits of the channel state in Eqn. \ref{eqn_6} are grouped into $2n$ qubit pairs and distributed among the participants such that each user receives one pair from each half of the channel in a cyclically shifted arrangement, i.e., the $I^{th}$ participant receives qubit pairs $(q_{2i-1},q_{2i})$ and $(q_{2(n+i)-3},q_{2(n+i)-2})$.

With this allocation, the protocol proceeds as follows:

\vspace{5pt}
{\justifying
\textit{Step 1}. Each participant performs two Bell-state measurements on pairs of message and resource qubits and communicates the resulting four classical bits to the adjacent user.

\textit{Step 2}. Each neighbouring participant partitions the received bits into two pairs, applying Pauli-$X$ corrections according to the second pair and Pauli-$Z$ corrections according to the first.
}
\vspace{5pt}

\section{Noise analysis}
\label{sec_6}

\begin{figure}
    \centering
    \includegraphics[width=\linewidth]{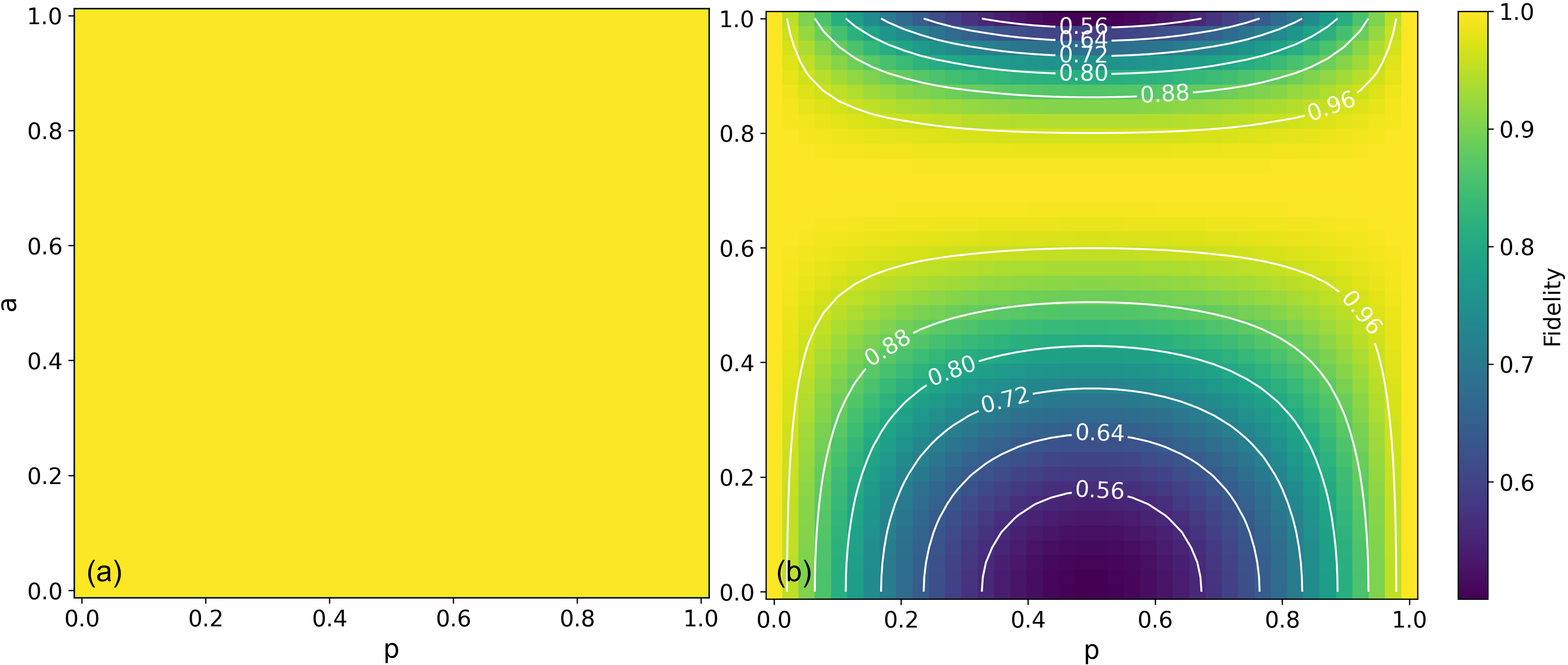}
    \caption{Fidelity as a function of noise parameter $p$ and input parameter $a$ under the influence of the Bit-flip noise channel for the (a) $3 \Leftrightarrow 3$ Bidirectional teleportation protocol and (b) $2 \Leftrightarrow 2 \Leftrightarrow 2$ Cyclic teleportation protocol. The value of parameters $b=c=\frac{1}{\sqrt2}$ is fixed for this analysis.}
    \label{fig_bf}
\end{figure}

\begin{figure}
    \centering
    \includegraphics[width=\linewidth]{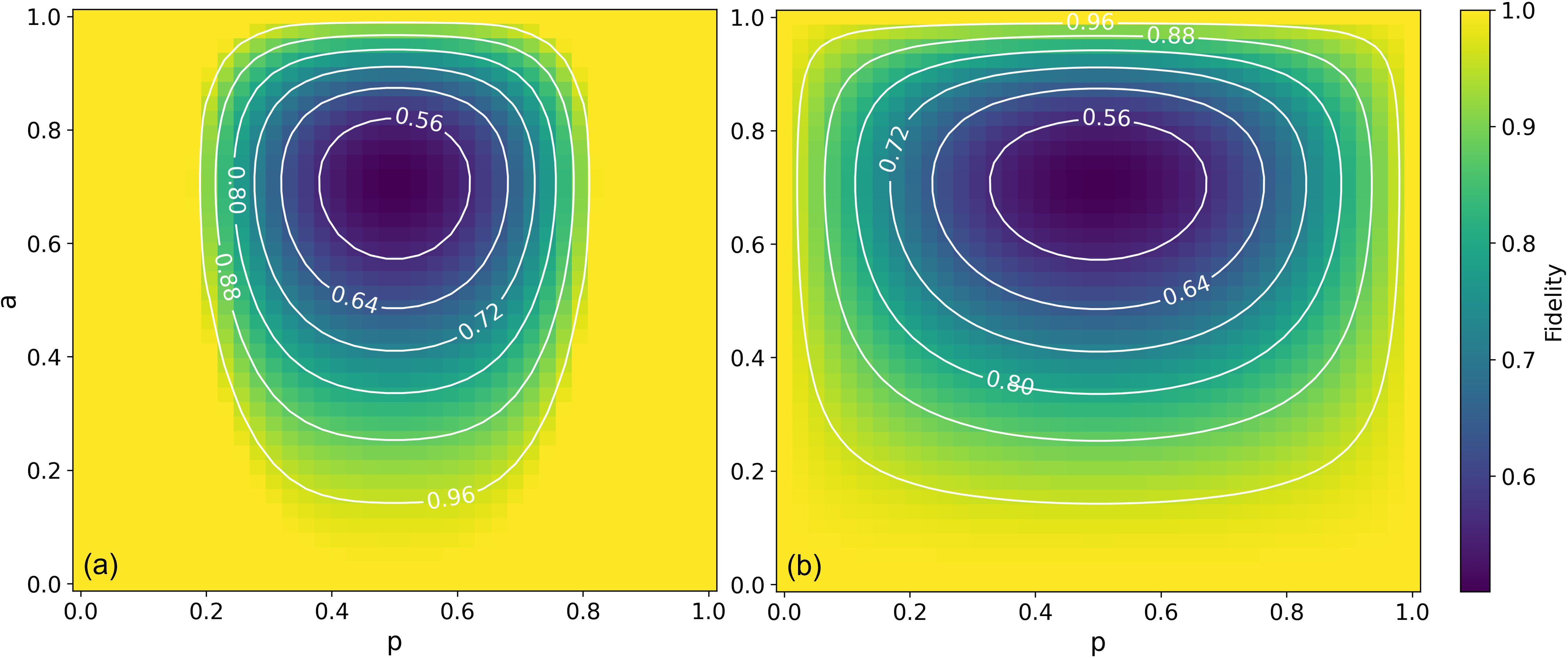}
    \caption{Fidelity as a function of noise parameter $p$ and input parameter $a$ under the influence of the Phase-flip noise channel for the (a) $3 \Leftrightarrow 3$ Bidirectional teleportation protocol and (b) $2 \Leftrightarrow 2 \Leftrightarrow 2$ Cyclic teleportation protocol. The value of parameters $b=c=\frac{1}{\sqrt2}$ is fixed for this analysis.}
    \label{fig_pf}
\end{figure}
\begin{figure}
    \centering
    \includegraphics[width=\linewidth]{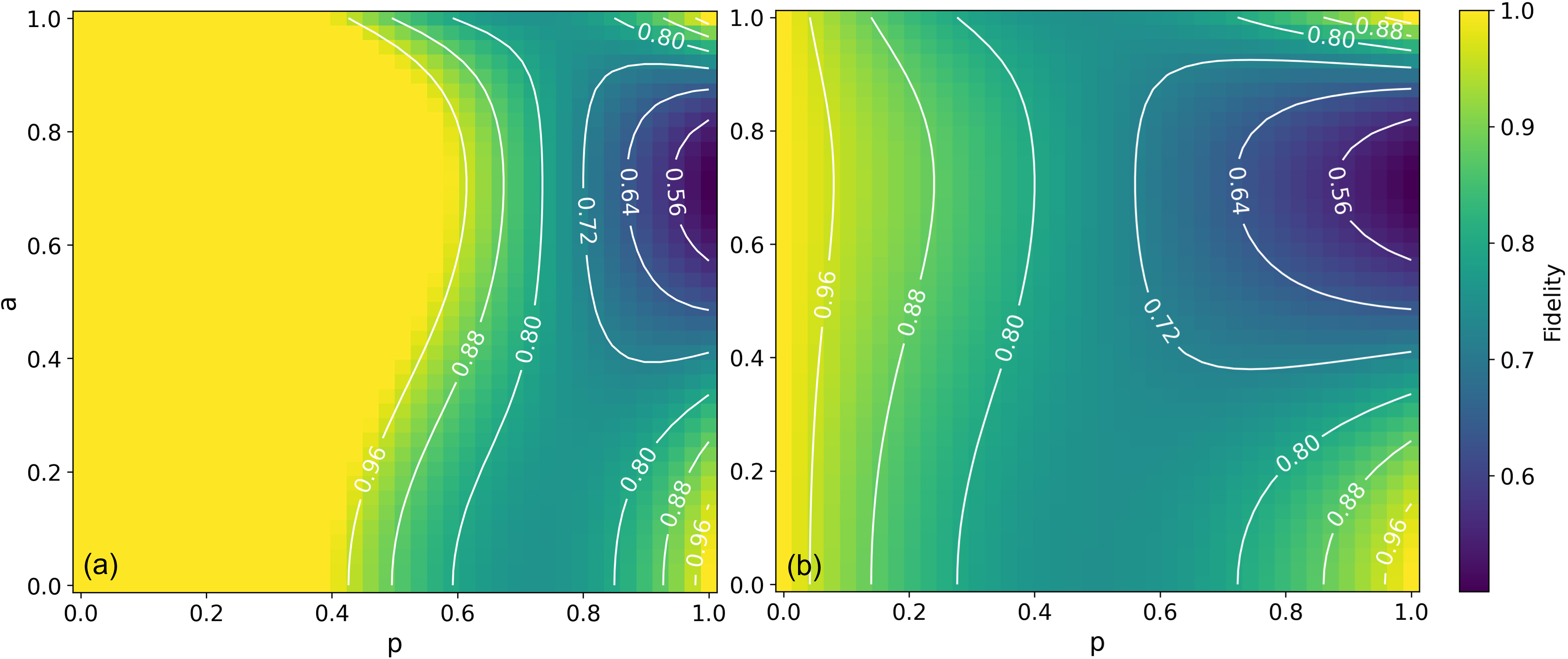}
    \caption{Fidelity as a function of noise parameter $p$ and input parameter $a$ under the influence of the Amplitude-damping noise channel for the (a) $3 \Leftrightarrow 3$ Bidirectional teleportation protocol and (b) $2 \Leftrightarrow 2 \Leftrightarrow 2$ Cyclic teleportation protocol. The value of parameters $b=c=\frac{1}{\sqrt2}$ is fixed for this analysis.}
    \label{fig_ad}
\end{figure}
\begin{figure}
    \centering
    \includegraphics[width=\linewidth]{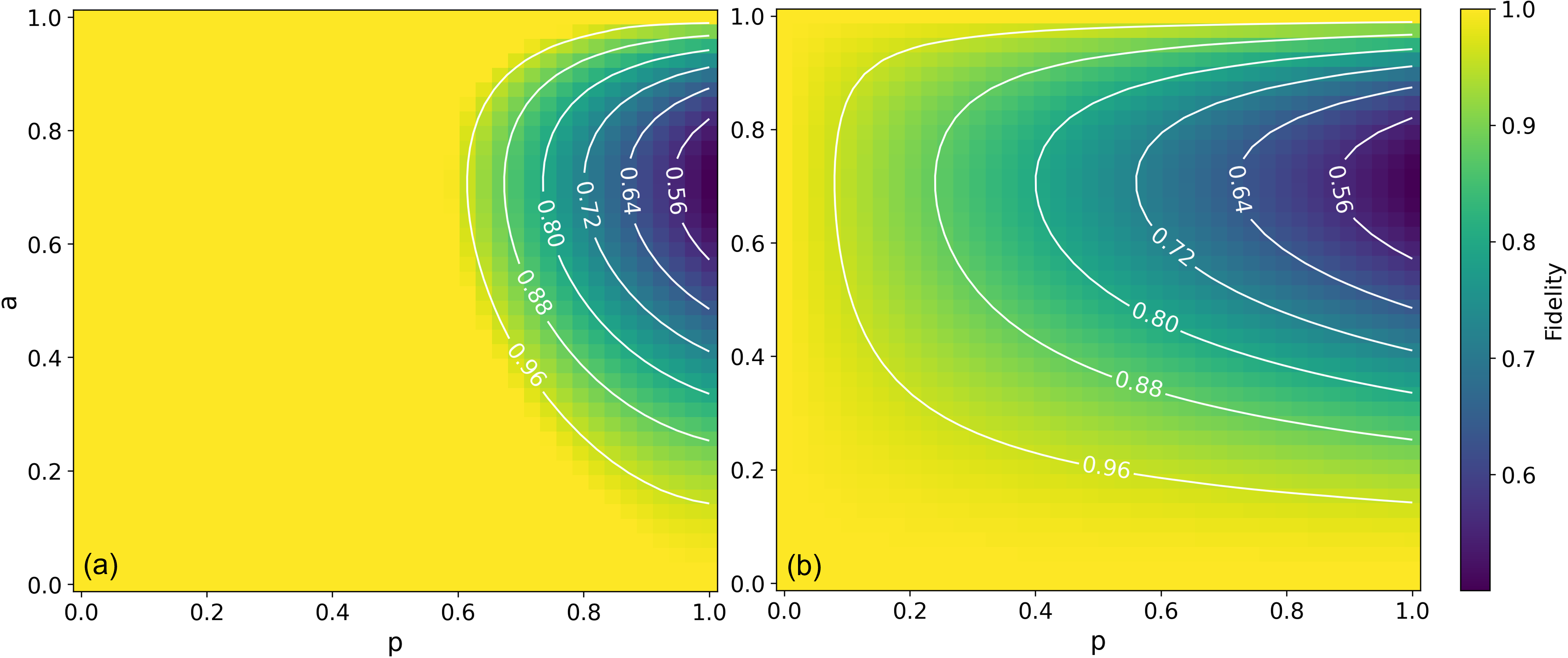}
    \caption{Fidelity as a function of noise parameter $p$ and input parameter $a$ under the influence of the Phase-damping noise channel for the (a) $3 \Leftrightarrow 3$ Bidirectional teleportation protocol and (b) $2 \Leftrightarrow 2 \Leftrightarrow 2$ Cyclic teleportation protocol. The value of parameters $b=c=\frac{1}{\sqrt2}$ is fixed for this analysis.}
    \label{fig_pd}
\end{figure}
\begin{figure}
    \centering
    \includegraphics[width=\linewidth]{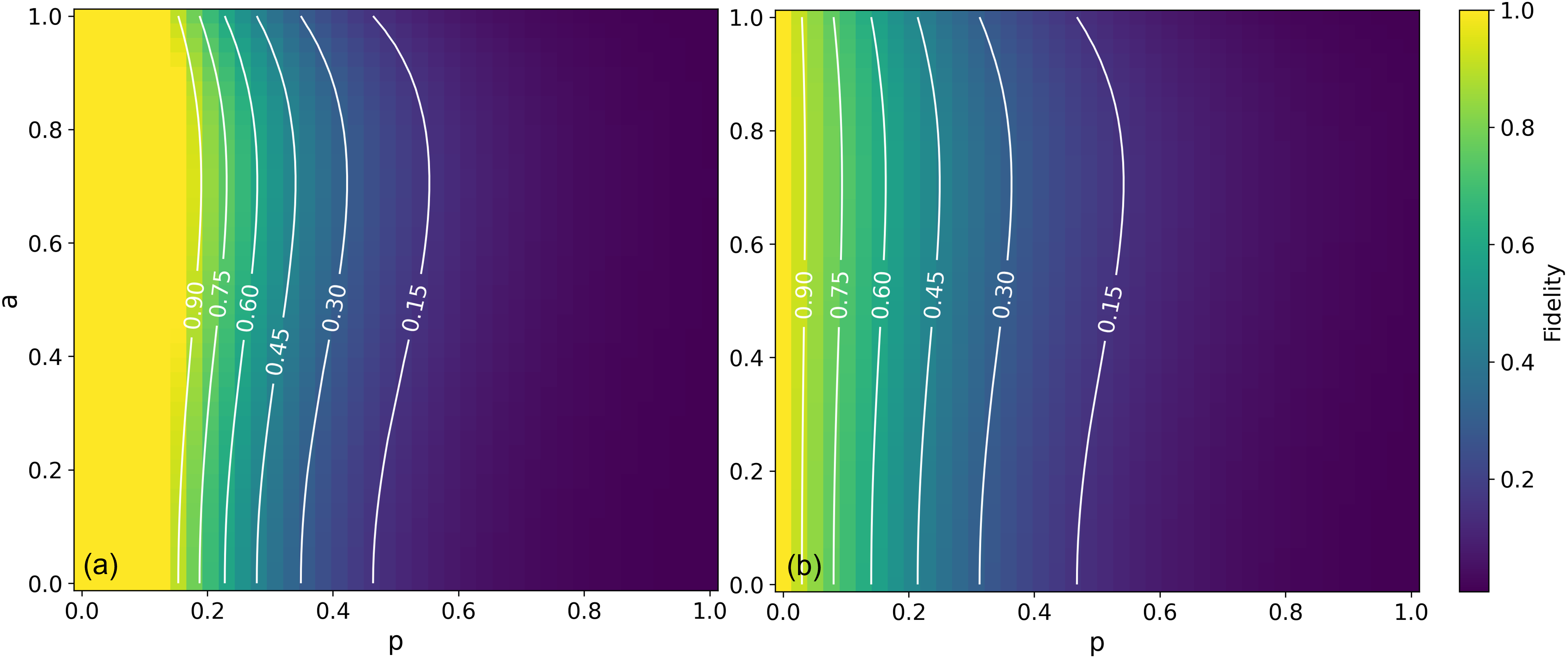}
    \caption{Fidelity as a function of noise parameter $p$ and input parameter $a$ under the influence of the Depolarizing noise channel for the (a) $3 \Leftrightarrow 3$ Bidirectional teleportation protocol and (b) $2 \Leftrightarrow 2 \Leftrightarrow 2$ Cyclic teleportation protocol. The value of parameters $b=c=\frac{1}{\sqrt2}$ is fixed for this analysis.}
    \label{fig_dp}
\end{figure}

\subsection{Fidelity of Teleportation}
In practical quantum communication, environmental interactions inevitably introduce noise that can affect the message states, resource qubits, and protocol operations. In this work, we focus exclusively on noise acting on the quantum channel qubits, since the entangled resource is particularly susceptible to decoherence during distribution, storage, and transmission prior to protocol execution. To evaluate protocol performance under realistic conditions, we analyze the teleportation fidelity of the bidirectional and cyclic schemes presented in Secs. \ref{sec_1} and \ref{sec_2} under five decoherence channels: Bit-Flip (BF), Phase-Flip (PF), Amplitude-Damping (AD), Phase-Damping (PD), and Depolarizing (DP) noise, modeled using Kraus operators in the operator-sum formalism. Fidelity quantifies the similarity between the sender's initial state and the receiver’s reconstructed state after corrective operations. The teleportation fidelity can be calculated as follows:
\begin{equation}
    F= \langle\Psi_{in}|\rho_{out}|\Psi_{in}\rangle
\end{equation}
where $|\Psi_{in}\rangle$ is the unknown input state, and $\rho_{out}$ is the noisy output density matrix obtained at the receiver's end. The noisy quantum channel can be written with the help of Kraus noise matrices in operator-sum representation as follows:
\begin{equation}
    \varepsilon(\rho_{12}) = \sum K |G_{12}\rangle_{1-12} \langle G_{12}| K^\dagger \nonumber
\end{equation}
\begin{align}
    \Rightarrow \sum_i (K_{i_{1}}^\mathbf{n}) \otimes (K_{i_{2}}^\mathbf{n}) \otimes (K_{i_{3}}^\mathbf{n}) \otimes (K_{i_{4}}^\mathbf{n})\otimes (K_{i_{5}}^\mathbf{n}) \otimes (K_{i_{6}}^\mathbf{n})  \otimes (K_{i_{7}}^\mathbf{n}) \otimes (K_{i_{8}}^\mathbf{n}) \otimes (K_{i_{9}}^\mathbf{n}) \nonumber \\
    \otimes (K_{i_{10}}^\mathbf{n}) \otimes (K_{i_{11}}^\mathbf{n}) \otimes (K_{i_{12}}^\mathbf{n}) \Big] \cdot \rho_{12} \cdot \Big[ (K_{i_{12}}^\mathbf{n})^\dagger \otimes (K_{i_{11}}^\mathbf{n})^\dagger \otimes (K_{i_{10}}^\mathbf{n})^\dagger \otimes (K_{i_{9}}^\mathbf{n})^\dagger \nonumber \\[5pt]
    \otimes (K_{i_{8}}^\mathbf{n})^\dagger \otimes (K_{i_{7}}^\mathbf{n})^\dagger\otimes (K_{i_{6}}^\mathbf{n})^\dagger \otimes (K_{i_{5}}^\mathbf{n})^\dagger  \otimes (K_{i_{4}}^\mathbf{n})^\dagger \otimes (K_{i_{3}}^\mathbf{n})^\dagger \otimes (K_{i_{2}}^\mathbf{n})^\dagger \otimes (K_{i_{1}}^\mathbf{n})^\dagger
\end{align}
where $\rho_{12} = |G_{12}\rangle\langle G_{12}|$ denotes the pure-state density matrix of resource state $|G_{12}\rangle$ in Eqn. \ref{eqn_1}. The terms $K_{i_{j}}^\mathbf{n}$ denotes Kraus operators of the noise model represented by $\mathbf{n}$, acting on the $j^{th}$ qubit, with $\mathbf{n} \in \{\text{BF, PF, AD, PD, DP}\}$ and $i=0,1$ for BF, PF, AD and PD noises, and $i=0,1,2,3$ for DP noise, and $\varepsilon(\rho_{12})$ is the non-unitary evolution of the quantum channel under noise. For the case of the bidirectional protocol in Sec. \ref{sec_1}, the output state $\rho^\text{Bi}_{out}$ is given as:
\begin{align}
    \rho_{out}^\text{Bi} = U_\text{Bi} \bigg[\mathrm{Tr}_{\substack{a_1,a_2,a_3,1,2,3\\b_1,b_2,b_3,4,5,6}} \Big\{\mathcal{E}_{a_1,1} \otimes \mathcal{E}_{a_2,2} \otimes \mathcal{E}_{a_3,3} \otimes \mathcal{E}_{b_1,4} \otimes \mathcal{E}_{b_2,5}\otimes \mathcal{E}_{b_3,6} \nonumber \\[-4.5pt]
    \cdot \big(\rho_{a_1 a_2 a_3}\otimes \rho_{b_1b_2b_3}\otimes\rho_{12}\big) \Big\} \bigg] U_\text{Bi}^\dagger
\end{align}
where $\rho_{a_1a_2a_3}$ and $\rho_{b_1b_2b_3}$ are density matrices corresponding to the message states of Alice and Bob respectively, $\mathcal{E}_{i,j}$ represents a Bell measurement on qubits $i$ and $j$, and $\mathrm{Tr}_{a_1-a_6}$ denotes the partial trace over all qubits subjected to BSM. The operators $U_\text{Bi}$ and its conjugate $U_\text{Bi}^\dagger$ correspond to the unitary corrections applied by the participants by using the measurement outcomes.
In a similar manner, the output state $\rho_{out}^\text{Cy}$ for the cyclic protocol described in Sec. \ref{sec_2} takes the form:
\begin{align}
    \rho_{out}^\text{Cy} = U_\text{Cy} \bigg[\mathrm{Tr}_{\substack{a_1,a_2,1,2\\b_1,b_2,3,4\\c_1,c_2,5,6}} \Big\{\mathcal{E}_{a_1,1} \otimes \mathcal{E}_{a_2,2} \otimes \mathcal{E}_{b_1,3} \otimes \mathcal{E}_{b_2,4} \otimes \mathcal{E}_{c_1,5}\otimes \mathcal{E}_{c_2,6} \nonumber \\[-9.5pt]
    \cdot\big(\rho_{a_1 a_2}\otimes \rho_{b_1b_2}\otimes\rho_{c_1c_2}\otimes\rho_{12}\big) \Big\} \bigg] U_\text{Cy}^\dagger
\end{align}
where $\rho_{a_1a_2}$, $\rho_{b_1b_2}$ and $\rho_{c_1c_2}$ are density matrices corresponding to the message states of Alice, Bob and Charlie respectively.

\subsection{Parameterization of message states}
In noisy implementations, teleportation fidelity depends on the transmitted message state. To investigate this dependence, the message states are parameterized by an input parameter while preserving normalization. For the bidirectional protocol in Sec. \ref{sec_1}, Alice's three-qubit message state in Eqn. \ref{eqn_3} is parametrized to:
\begin{align}
    |\Psi_A\rangle_{a_1a_2a_3} = \frac{1}{2} \Big(a|000\rangle + \sqrt{1-a^2}|001\rangle + a|010\rangle + \sqrt{1-a^2}|011\rangle& \nonumber \\ 
    + \sqrt{1-a^2}|100\rangle + a|101\rangle + \sqrt{1-a^2}|110\rangle + a|111\rangle& \Big)
\end{align}
where $a\in[0,1]$. Bob's state is parametrized in a similar fashion, where his input parameter is represented by $b$. We then vary $a$ in the interval $[0,1]$ such that the input state transitions continuously while preserving its normalization at all times. Especially of note are three $a$ values: the extremal values of $a=0$ and $a=1$, and the intermediate value of $a=\frac{1}{\sqrt2}$.
Similarly for the cyclic protocol in Sec. \ref{sec_2}, the two-qubit message states of Eqn. \ref{eqn_7} are parametrized in the following way:
\begin{equation}
    |\Psi^{(2)}_A\rangle_{a_1a_2} = \frac{1}{\sqrt2}(a|00\rangle + \sqrt{1-a^2}|01\rangle + \sqrt{1-a^2}|10\rangle + a|11\rangle)
\end{equation}
This framework enables an analytical evaluation of the teleportation fidelity as a function of the input parameter $a$.

\subsection{Variation of Fidelity}






The fidelity variation under the studied noise channels are given in Fig. \ref{fig_bf} for Bit-Flip (BF), Fig. \ref{fig_pf} for Phase-Flip (PF), Fig. \ref{fig_ad} for Amplitude-Damping (AD), Fig. \ref{fig_pd} for Phase-Damping (PD) and Fig. \ref{fig_dp} for Depolarizing (DP) noises respectively. Across all noise models, the teleportation fidelity of both protocols is governed by the underlying physical action of the channel on the transmitted quantum information. Flip-type noises exhibit characteristic symmetry with respect to the noise strength, yielding unit fidelity at both $p=0$ and $p=1$, whereas dissipative and decohering channels produce irreversible degradation with increasing $p$. Among the flip-type channels, bit-flip noise constitutes a special case: the bidirectional protocol remains completely immune, maintaining unit fidelity irrespective of the input state and noise strength, while the cyclic protocol exhibits a symmetric fidelity variation in which basis states ($a=0,1$) degrade to a minimum at $p=\frac{1}{2}$ before recovering to unity at $p=1$ owing to the restoration of the channel correlations by multiple bit-flip errors, whereas equal-superposition states ($a=\frac{1}{\sqrt2}$) remain unaffected throughout. In contrast, phase-flip noise primarily degrades phase coherence, leaving basis states largely invariant while causing the strongest fidelity reduction for equal-superposition states, with both protocols displaying symmetric dependence on $p$, although the cyclic protocol experiences a more pronounced decline. A similar preference for basis states is observed under amplitude- and phase-damping channels, where the irreversible loss of excitation and coherence causes a monotonic decrease in fidelity, most severely affecting superposition states and leaving basis-like states comparatively robust. Depolarizing noise differs from the preceding cases by affecting all state components almost uniformly, producing a rapid fidelity reduction with only weak dependence on the input-state parameter as the transmitted state approaches a maximally mixed state. Overall, while the qualitative dependence of fidelity on the input-state structure remains consistent across both protocols for each noise model, the cyclic protocol invariably exhibits a steeper degradation than the bidirectional protocol under all non-trivial noise channels, demonstrating its comparatively higher sensitivity to environmental disturbances.

\begin{table}[b]
    \caption{\label{tab_1}Comparison of our bidirectional protocol with protocols from the literature.}
    \begin{tabular}{ l l l l l l l }
        \toprule
        \textbf{References} & \textbf{Type} & \textbf{QI}& \textbf{QR}& \textbf{CB} & \textbf{NA} & \textbf{EF $(\%)$} \\
        \midrule
        Li et. al.\cite{li_bidirectional_2016} & CBQT & 4 & 9 & 9 & \checkmark & 22  \\
        Zhou et. al.\cite{zhou_asymmetric_2019} & CABQT & 4 & 9 & 9 & \checkmark & 22 \\
        Huo et. al.\cite{huo_controlled_2021} & CABQT & 5 & 11 &  11  & $\times$ & 22.7  \\
        Jiang et. al.\cite{jiang_bidirectional_2021} & CBQT & 6 & 11 & 11 & $\times$ & 27.3 \\
        Dai et. al.\cite{dai_asymmetric_2022} & ABQT & 5 & 9 & 9 & $\times$ & 27.8  \\
        Wang et. al.\cite{wang_bidirectional_2022} & CBQT & 4 & 10 & 10 & \checkmark & 20  \\
        Choudhury et. al.\cite{choudhury_bidirectional_2023} & BQT & 2 & 5 & 4 & $\times$ & 22.2 \\
        \midrule
        \textbf{Our BQT protocol} & \textbf{BQT} & \textbf{6} & \textbf{12} & \textbf{12} & \checkmark & \textbf{25} \\
        \botrule
    \end{tabular}
    \begin{tablenotes}
        \footnotesize
        \item \textit{BQT}: Bidirectional Quantum Teleportation,
        \item \textit{ABQT}: Asymmetric Bidirectional Quantum Teleportation, 
        \item \textit{CBQT}: Controlled Bidirectional Quantum Teleportation,
        \item \textit{CABQT}: Controlled Asymmetric Bidirectional Quantum Teleportation.
    \end{tablenotes}
\end{table}

\section{\label{sec_7}Intrinsic Efficiency}
The intrinsic efficiency $\eta$ quantifies how effectively a quantum teleportation protocol utilizes its quantum and classical resources for information transfer. It is defined as the ratio of successfully transmitted message qubits to the total resources consumed during the protocol. Thus, higher values of $\eta$ correspond to more efficient resource utilization, particularly for deterministic teleportation schemes.
It is typically defined as \cite{yuan_optimizing_2008}: 
\begin{equation}
\label{eqn_2}
    \eta = \frac{q_s}{q_u+b_t} \quad \Rightarrow \quad \eta = \frac{6}{12 +12 } = 25 \%.
\end{equation}
where $q_s$ denotes the number of message qubits representing the quantum information transferred, channel qubits $q_u$ forming the quantum resource state for the protocol and $b_t$ is the total classical bits exchanged between the parties to complete the reconstruction of all message states.
The intrinsic efficiencies of the proposed two-party bidirectional scheme in Sec. \ref{sec_1} and the three-party cyclic scheme in Sec. \ref{sec_2} are calculated to be $25\%$. 

Table \ref{tab_1} provides a comparative analysis of the proposed bidirectional protocol against existing schemes in the literature, while Table \ref{tab_2} presents a corresponding comparison for the proposed cyclic protocol with related works. The comparisons are based on several key parameters, including the protocol type and configuration (Type), the total number of quantum information qubits being transmitted (QI), the number of channel qubits constituting the resource (QR), the number of classical bits obtained as a result of Bell-state measurements by the participants (CB), the inclusion of noise analysis under different decoherence models (NA), and the obtained intrinsic efficiency of the protocol (EF).

\begin{table}
    \caption{\label{tab_2}Comparison of our cyclic protocol with protocols from the literature.}
        \begin{tabular}{ l l l l l l l }
            \toprule
            \textbf{References} & \textbf{Type} & \textbf{QI}& \textbf{QR}& \textbf{CB} & \textbf{NA} & \textbf{EF $(\%)$} \\
            \midrule
            Li et. al.\cite{li_tripartite_2016} & CCYQT & 3 & 7 & 7 & $\times$ & 21 \\
            Verma et. al.\cite{verma_improvement_2021} & CCYQT & 3 & 7 & 7 & $\times$ & 21.4 \\
            Zhou et. al.\cite{zhou_asymmetric_2021} & ACCYQT & 5 & 9 & 9 & \checkmark & 27.7 \\
            Rahmawati et. al.\cite{rahmawati_symmetric_2022} & CCYQT & 3 & 7 & 7 & $\times$ & 21 \\
            Mahjoory et. al.\cite{mahjoory_asymmetric_2023} & CCYQT & 6 & 12 &9 & $\times$ & 28.5 \\
            Jiang et. al.\cite{jiang_multi-party_2024} & CYBQT & 6 & 13 & 15 & $\times$ & 21.4 \\
            Taufiqi et. al.\cite{taufiqi_cyclic_2024} & ACCYQT & 6 & 13 & 13 & \checkmark & 23.1 \\
            Kaur et. al.\cite{kaur_asymmetric_2025} & CCYQT & 3 & 9 & 12 & $\times$ & 14.3 \\
            \midrule
            \textbf{Our CyQT protocol} & \textbf{CYQT} & \textbf{6} & \textbf{12} & \textbf{12} & \checkmark & \textbf{25} \\
            \botrule
        \end{tabular}
        \begin{tablenotes}
            \item \textit{ACCYQT}: Asymmetric Controlled Cyclic Quantum Teleportation,
            \item \textit{CCYQT}: Controlled Cyclic Quantum Teleportation.
        \end{tablenotes}
\end{table}

\section{\label{sec_8}Results and Discussion}
In this work, we successfully demonstrated a unified quantum teleportation framework capable of implementing both bidirectional and cyclic communication protocols using a single twelve-qubit entangled resource. Unlike conventional protocol-specific approaches, the proposed framework realizes distinct communication tasks through qubit allocation, Bell-state measurements, and classical communication without modifying the underlying quantum channel. The bidirectional protocol enables simultaneous exchange of arbitrary three-qubit states between two parties, while the cyclic protocol facilitates transfer of general two-qubit states among three users in a looped topology. Both schemes were further generalized to support scalable multi-qubit and multi-party quantum communication networks. An important feature of the framework is the recursive structure of the entangled resource. Following successive Bell-state measurements, the channel reduces progressively from $|G_{12}\rangle$ to lower-dimensional states such as $|G_{10}\rangle$ and $|G_8\rangle$, while preserving the same underlying entanglement structure. This structural invariance plays a key role in maintaining the scalability and operational consistency of the protocols.

The effects of amplitude-damping, phase-damping, bit-flip, phase-flip, and depolarizing noise channels on teleportation fidelity were analyzed as functions of both the message-state parameters and noise strength. The results show that fidelity strongly depends on the both the type of noise affecting the quantum channel and the structure of the message states of the protocol. Amplitude damping produced the strongest degradation due to irreversible energy loss, particularly for superposition states with larger excited-state contributions. Phase damping primarily suppressed coherence while preserving state populations, making basis states comparatively robust. Bit-flip and phase-flip channels exhibited symmetry-dependent fidelity behavior, whereas depolarizing noise caused an almost uniform degradation across all input states. Comparative analysis further revealed that the cyclic protocol is generally more sensitive to decoherence than the bidirectional scheme, especially under dissipative and depolarizing environments. The study also identified classes of input states that remain relatively resilient under specific noise models, suggesting possible strategies for noise-aware state preparation.

Finally, the intrinsic efficiency of the proposed protocols was evaluated and compared with existing schemes, demonstrating improved resource utilization through the reuse of a common entangled channel for multiple communication tasks. Overall, the proposed framework provides a scalable and adaptable approach to multitasking quantum communication, with potential applications in fault-tolerant quantum communication architectures.

\clearpage

\begin{appendices}

\begin{table}[b]
    \caption{\label{tab_3} Table for each sender-receiver pair, summarizing the receiver's Bell-state measurements, along with the corresponding receiver's post-measurement states and corrective unitary operations for the \textbf{bidirectional protocol} described in Sec. \ref{sec_1}.}
    \centering
    \begin{tabular}{l r l}
        \toprule
        \textbf{Results} & \textbf{Post-measurement State} & \textbf{Unitaries} \\
        \midrule
        
        $|\phi^+\rangle, |\phi^+\rangle, |\phi^+\rangle$ & 
$\mu_0|000\rangle + \mu_1|001\rangle + \mu_2|010\rangle +\mu_3|011\rangle + \mu_4|100\rangle + \mu_5|101\rangle + \mu_6|110\rangle + \mu_7|111\rangle$ & $I \otimes I \otimes I$ \\
        $|\phi^+\rangle, |\phi^+\rangle, |\psi^+\rangle$ & 
$\mu_1|000\rangle + \mu_0|001\rangle + \mu_3|010\rangle + \mu_2|011\rangle + \mu_5|100\rangle + \mu_4|101\rangle + \mu_7|110\rangle + \mu_6|111\rangle$ & $I \otimes I \otimes X$ \\
        $|\phi^+\rangle, |\phi^+\rangle, |\phi^-\rangle$ & 
$\mu_0|000\rangle - \mu_1|001\rangle + \mu_2|010\rangle -\mu_3|011\rangle + \mu_4|100\rangle - \mu_5|101\rangle + \mu_6|110\rangle - \mu_7|111\rangle$ & $I \otimes I \otimes Z$ \\
        $|\phi^+\rangle, |\phi^+\rangle, |\psi^-\rangle$ & 
$ - \mu_1|000\rangle + \mu_0|001\rangle - \mu_3|010\rangle + \mu_2|011\rangle - \mu_5|100\rangle + \mu_4|101\rangle - \mu_7|110\rangle + \mu_6|111\rangle$ & $I \otimes I \otimes ZX$ \\[2pt]

        $|\phi^+\rangle, |\psi^+\rangle, |\phi^+\rangle$ &
$\mu_2|000\rangle + \mu_3|001\rangle + \mu_0|010\rangle +\mu_1|011\rangle + \mu_6|100\rangle + \mu_7|101\rangle + \mu_4|110\rangle + \mu_5|111\rangle$ & $I \otimes X \otimes I$ \\
        $|\phi^+\rangle, |\psi^+\rangle, |\psi^+\rangle$ &
$\mu_3|000\rangle + \mu_2|001\rangle + \mu_1|010\rangle +\mu_0|011\rangle + \mu_7|100\rangle + \mu_6|101\rangle + \mu_5|110\rangle + \mu_4|111\rangle$ & $I \otimes X \otimes X$ \\
        $|\phi^+\rangle, |\psi^+\rangle, |\phi^-\rangle$ &
$\mu_2|000\rangle - \mu_3|001\rangle + \mu_0|010\rangle - \mu_1|011\rangle + \mu_6|100\rangle - \mu_7|101\rangle + \mu_4|110\rangle - \mu_5|111\rangle$ & $I \otimes X \otimes Z$ \\
        $|\phi^+\rangle, |\psi^+\rangle, |\psi^-\rangle$ &
$-\mu_3|000\rangle + \mu_2|001\rangle - \mu_1|010\rangle + \mu_0|011\rangle - \mu_7|100\rangle + \mu_6|101\rangle - \mu_5|110\rangle + \mu_4|111\rangle$ & $I \otimes X \otimes ZX$ \\[2pt]
        
        $|\phi^+\rangle, |\phi^-\rangle, |\phi^+\rangle$ & 
$\mu_0|000\rangle + \mu_1|001\rangle - \mu_2|010\rangle - \mu_3|011\rangle + \mu_4|100\rangle + \mu_5|101\rangle - \mu_6|110\rangle - \mu_7|111\rangle$ & $I \otimes Z \otimes I$ \\
        $|\phi^+\rangle, |\phi^-\rangle, |\psi^+\rangle$ &
$\mu_1|000\rangle + \mu_0|001\rangle - \mu_3|010\rangle - \mu_2|011\rangle + \mu_5|100\rangle + \mu_4|101\rangle - \mu_7|110\rangle - \mu_6|111\rangle$ & $I \otimes Z \otimes X$ \\
        $|\phi^+\rangle, |\phi^-\rangle, |\phi^-\rangle$ &
$\mu_0|000\rangle - \mu_1|001\rangle - \mu_2|010\rangle + \mu_3|011\rangle + \mu_4|100\rangle - \mu_5|101\rangle - \mu_6|110\rangle + \mu_7|111\rangle$ & $I \otimes Z \otimes Z$ \\
        $|\phi^+\rangle, |\phi^-\rangle, |\psi^-\rangle$ &
$-\mu_1|000\rangle + \mu_0|001\rangle + \mu_3|010\rangle - \mu_2|011\rangle - \mu_5|100\rangle + \mu_4|101\rangle + \mu_7|110\rangle - \mu_6|111\rangle$ & $I \otimes Z \otimes ZX$ \\[2pt]
        
        $|\phi^+\rangle, |\psi^-\rangle, |\phi^+\rangle$ &
$-\mu_2|000\rangle - \mu_3|001\rangle + \mu_0|010\rangle + \mu_1|011\rangle - \mu_6|100\rangle - \mu_7|101\rangle + \mu_4|110\rangle + \mu_5|111\rangle$ & $I \otimes ZX \otimes I$ \\
        $|\phi^+\rangle, |\psi^-\rangle, |\psi^+\rangle$ &
$-\mu_3|000\rangle - \mu_2|001\rangle + \mu_1|010\rangle + \mu_0|011\rangle - \mu_7|100\rangle - \mu_6|101\rangle + \mu_5|110\rangle + \mu_4|111\rangle$ & $I \otimes ZX \otimes X$ \\
        $|\phi^+\rangle, |\psi^-\rangle, |\phi^-\rangle$ &
$-\mu_2|000\rangle + \mu_3|001\rangle + \mu_0|010\rangle - \mu_1|011\rangle - \mu_6|100\rangle + \mu_7|101\rangle + \mu_4|110\rangle - \mu_5|111\rangle$ & $I \otimes ZX \otimes Z$ \\
        $|\phi^+\rangle, |\psi^-\rangle, |\psi^-\rangle$ &
$\mu_3|000\rangle - \mu_2|001\rangle - \mu_1|010\rangle + \mu_0|011\rangle + \mu_7|100\rangle - \mu_6|101\rangle - \mu_5|110\rangle + \mu_4|111\rangle$ & $I \otimes ZX \otimes ZX$ \\[2pt]

        $|\psi^+\rangle, |\phi^+\rangle, |\phi^+\rangle$ & 
$\mu_4|000\rangle + \mu_5|001\rangle + \mu_6|010\rangle + \mu_7|011\rangle + \mu_0|100\rangle + \mu_1|101\rangle + \mu_2|110\rangle + \mu_3|111\rangle$ & $X \otimes I \otimes I$ \\
        $|\psi^+\rangle, |\phi^+\rangle, |\psi^+\rangle$ &
$\mu_5|000\rangle + \mu_4|001\rangle + \mu_7|010\rangle + \mu_6|011\rangle + \mu_1|100\rangle + \mu_0|101\rangle + \mu_3|110\rangle + \mu_2|111\rangle$ & $X \otimes I \otimes X$ \\
        $|\psi^+\rangle, |\phi^+\rangle, |\phi^-\rangle$ &
$\mu_4|000\rangle - \mu_5|001\rangle + \mu_6|010\rangle - \mu_7|011\rangle + \mu_0|100\rangle - \mu_1|101\rangle + \mu_2|110\rangle - \mu_3|111\rangle$ & $X \otimes I \otimes Z$ \\
        $|\psi^+\rangle, |\phi^+\rangle, |\psi^-\rangle$ & 
$-\mu_5|000\rangle + \mu_4|001\rangle - \mu_7|010\rangle + \mu_6|011\rangle - \mu_1|100\rangle + \mu_0|101\rangle - \mu_3|110\rangle + \mu_2|111\rangle$ & $X \otimes I \otimes ZX$ \\[2pt]
        
        $|\psi^+\rangle, |\psi^+\rangle, |\phi^+\rangle$ & 
$\mu_6|000\rangle + \mu_7|001\rangle + \mu_4|010\rangle + \mu_5|011\rangle + \mu_2|100\rangle + \mu_3|101\rangle + \mu_0|110\rangle + \mu_1|111\rangle$ & $X \otimes X \otimes I$ \\
        $|\psi^+\rangle, |\psi^+\rangle, |\psi^+\rangle$ &
$\mu_7|000\rangle + \mu_6|001\rangle + \mu_5|010\rangle + \mu_4|011\rangle + \mu_3|100\rangle + \mu_2|101\rangle + \mu_1|110\rangle + \mu_0|111\rangle$ & $X \otimes X \otimes X$ \\
        $|\psi^+\rangle, |\psi^+\rangle, |\phi^-\rangle$ &
$\mu_6|000\rangle - \mu_7|001\rangle + \mu_4|010\rangle - \mu_5|011\rangle + \mu_2|100\rangle - \mu_3|101\rangle + \mu_0|110\rangle - \mu_1|111\rangle$ & $X \otimes X \otimes Z$ \\
        $|\psi^+\rangle, |\psi^+\rangle, |\psi^-\rangle$ &
$-\mu_7|000\rangle + \mu_6|001\rangle - \mu_5|010\rangle + \mu_4|011\rangle - \mu_3|100\rangle + \mu_2|101\rangle - \mu_1|110\rangle + \mu_0|111\rangle$ & $X \otimes X \otimes ZX$ \\[2pt]

        $|\psi^+\rangle, |\phi^-\rangle, |\phi^+\rangle$ &
$\mu_4|000\rangle + \mu_5|001\rangle - \mu_6|010\rangle - \mu_7|011\rangle + \mu_0|100\rangle + \mu_1|101\rangle - \mu_2|110\rangle - \mu_3|111\rangle$ & $X \otimes Z \otimes I$ \\
        $|\psi^+\rangle, |\phi^-\rangle, |\psi^+\rangle$ & 
$\mu_5|000\rangle + \mu_4|001\rangle - \mu_7|010\rangle - \mu_6|011\rangle + \mu_1|100\rangle + \mu_0|101\rangle - \mu_3|110\rangle - \mu_2|111\rangle$ & $X \otimes Z \otimes X$ \\
        $|\psi^+\rangle, |\phi^-\rangle, |\phi^-\rangle$ &
$\mu_4|000\rangle - \mu_5|001\rangle - \mu_6|010\rangle + \mu_7|011\rangle + \mu_0|100\rangle - \mu_1|101\rangle - \mu_2|110\rangle + \mu_3|111\rangle$ & $X \otimes Z \otimes Z$ \\
        $|\psi^+\rangle, |\phi^-\rangle, |\psi^-\rangle$ &
$-\mu_5|000\rangle + \mu_4|001\rangle + \mu_7|010\rangle - \mu_6|011\rangle - \mu_1|100\rangle + \mu_0|101\rangle + \mu_3|110\rangle - \mu_2|111\rangle$ & $X \otimes Z \otimes ZX$ \\[2pt]
        
        $|\psi^+\rangle, |\psi^-\rangle, |\phi^+\rangle$ & 
$-\mu_6|000\rangle - \mu_7|001\rangle + \mu_4|010\rangle + \mu_5|011\rangle - \mu_2|100\rangle - \mu_3|101\rangle + \mu_0|110\rangle + \mu_1|111\rangle$ & $X \otimes ZX \otimes I$ \\
        $|\psi^+\rangle, |\psi^-\rangle, |\psi^+\rangle$ &
$-\mu_7|000\rangle - \mu_6|001\rangle + \mu_5|010\rangle + \mu_4|011\rangle - \mu_3|100\rangle - \mu_2|101\rangle + \mu_1|110\rangle + \mu_0|111\rangle$ & $X \otimes ZX \otimes X$ \\
        $|\psi^+\rangle, |\psi^-\rangle, |\phi^-\rangle$ &
$-\mu_6|000\rangle + \mu_7|001\rangle + \mu_4|010\rangle - \mu_5|011\rangle - \mu_2|100\rangle + \mu_3|101\rangle + \mu_0|110\rangle - \mu_1|111\rangle$ & $X \otimes ZX \otimes Z$ \\
        $|\psi^+\rangle, |\psi^-\rangle, |\psi^-\rangle$ &
$\mu_7|000\rangle - \mu_6|001\rangle - \mu_5|010\rangle + \mu_4|011\rangle + \mu_3|100\rangle - \mu_2|101\rangle - \mu_1|110\rangle + \mu_0|111\rangle$ & $X \otimes ZX \otimes ZX$ \\[2pt]

        $|\phi^-\rangle, |\phi^+\rangle, |\phi^+\rangle$ & 
$\mu_0|000\rangle + \mu_1|001\rangle + \mu_2|010\rangle + \mu_3|011\rangle - \mu_4|100\rangle - \mu_5|101\rangle - \mu_6|110\rangle - \mu_7|111\rangle$ & $Z \otimes I \otimes I$ \\
        $|\phi^-\rangle, |\phi^+\rangle, |\psi^+\rangle$ & 
$\mu_1|000\rangle + \mu_0|001\rangle + \mu_3|010\rangle + \mu_2|011\rangle - \mu_5|100\rangle - \mu_4|101\rangle - \mu_7|110\rangle - \mu_6|111\rangle$ & $Z \otimes I \otimes X$ \\
        $|\phi^-\rangle, |\phi^+\rangle, |\phi^-\rangle$ & 
$\mu_0|000\rangle - \mu_1|001\rangle + \mu_2|010\rangle - \mu_3|011\rangle - \mu_4|100\rangle + \mu_5|101\rangle - \mu_6|110\rangle + \mu_7|111\rangle$ & $Z \otimes I \otimes Z$ \\
        $|\phi^-\rangle, |\phi^+\rangle, |\psi^-\rangle$ & 
$-\mu_1|000\rangle + \mu_0|001\rangle - \mu_3|010\rangle + \mu_2|011\rangle + \mu_5|100\rangle - \mu_4|101\rangle + \mu_7|110\rangle - \mu_6|111\rangle$ & $Z \otimes I \otimes ZX$ \\[2pt]

        $|\phi^-\rangle, |\psi^+\rangle, |\phi^+\rangle$ &
$\mu_2|000\rangle + \mu_3|001\rangle + \mu_0|010\rangle + \mu_1|011\rangle - \mu_6|100\rangle - \mu_7|101\rangle - \mu_4|110\rangle - \mu_5|111\rangle$ & $Z \otimes X \otimes I$ \\
        $|\phi^-\rangle, |\psi^+\rangle, |\psi^+\rangle$ & 
$\mu_3|000\rangle + \mu_2|001\rangle + \mu_1|010\rangle + \mu_0|011\rangle - \mu_7|100\rangle - \mu_6|101\rangle - \mu_5|110\rangle - \mu_4|111\rangle$ & $Z \otimes X \otimes X$ \\
        $|\phi^-\rangle, |\psi^+\rangle, |\phi^-\rangle$ & 
$\mu_2|000\rangle - \mu_3|001\rangle + \mu_0|010\rangle - \mu_1|011\rangle - \mu_6|100\rangle + \mu_7|101\rangle - \mu_4|110\rangle + \mu_5|111\rangle$ & $Z \otimes X \otimes Z$ \\
        $|\phi^-\rangle, |\psi^+\rangle, |\psi^-\rangle$ & 
$-\mu_3|000\rangle + \mu_2|001\rangle - \mu_1|010\rangle + \mu_0|011\rangle + \mu_7|100\rangle - \mu_6|101\rangle + \mu_5|110\rangle - \mu_4|111\rangle$ & $Z \otimes X \otimes ZX$ \\[2pt]

        \botrule
        \multicolumn{3}{l}{Continued on next page} \\
    \end{tabular}
\end{table}

\begin{table}
    \caption*{Table \ref{tab_3}. (Continued)}
    \centering
    \begin{tabular}{l r l}
        \toprule
        \textbf{Results} & \textbf{Post-measurement State} & \textbf{Unitaries} \\
        \midrule

        $|\phi^-\rangle, |\phi^-\rangle, |\phi^+\rangle$ & 
$\mu_0|000\rangle + \mu_1|001\rangle - \mu_2|010\rangle - \mu_3|011\rangle - \mu_4|100\rangle - \mu_5|101\rangle + \mu_6|110\rangle + \mu_7|111\rangle$ & $Z \otimes Z \otimes I$ \\
        $|\phi^-\rangle, |\phi^-\rangle, |\psi^+\rangle$ & 
$\mu_1|000\rangle + \mu_0|001\rangle - \mu_3|010\rangle - \mu_2|011\rangle - \mu_5|100\rangle - \mu_4|101\rangle + \mu_7|110\rangle + \mu_6|111\rangle$ & $Z \otimes Z \otimes X$ \\
        $|\phi^-\rangle, |\phi^-\rangle, |\phi^-\rangle$ & 
$\mu_0|000\rangle - \mu_1|001\rangle - \mu_2|010\rangle + \mu_3|011\rangle - \mu_4|100\rangle + \mu_5|101\rangle + \mu_6|110\rangle - \mu_7|111\rangle$ & $Z \otimes Z \otimes Z$ \\
        $|\phi^-\rangle, |\phi^-\rangle, |\psi^-\rangle$ & 
$-\mu_1|000\rangle + \mu_0|001\rangle + \mu_3|010\rangle - \mu_2|011\rangle + \mu_5|100\rangle - \mu_4|101\rangle - \mu_7|110\rangle + \mu_6|111\rangle$ & $Z \otimes Z \otimes ZX$ \\[2pt]
        
        $|\phi^-\rangle, |\psi^-\rangle, |\phi^+\rangle$ & 
$-\mu_2|000\rangle - \mu_3|001\rangle + \mu_0|010\rangle + \mu_1|011\rangle + \mu_6|100\rangle + \mu_7|101\rangle - \mu_4|110\rangle - \mu_5|111\rangle$ & $Z \otimes ZX \otimes I$ \\
        $|\phi^-\rangle, |\psi^-\rangle, |\psi^+\rangle$ & 
$-\mu_3|000\rangle - \mu_2|001\rangle + \mu_1|010\rangle + \mu_0|011\rangle + \mu_7|100\rangle + \mu_6|101\rangle - \mu_5|110\rangle - \mu_4|111\rangle$ & $Z \otimes ZX \otimes X$ \\
        $|\phi^-\rangle, |\psi^-\rangle, |\phi^-\rangle$ & 
$-\mu_2|000\rangle + \mu_3|001\rangle + \mu_0|010\rangle - \mu_1|011\rangle + \mu_6|100\rangle - \mu_7|101\rangle - \mu_4|110\rangle + \mu_5|111\rangle$ & $Z \otimes ZX \otimes Z$ \\
        $|\phi^-\rangle, |\psi^-\rangle, |\psi^-\rangle$ & 
$\mu_3|000\rangle - \mu_2|001\rangle - \mu_1|010\rangle + \mu_0|011\rangle - \mu_7|100\rangle + \mu_6|101\rangle + \mu_5|110\rangle - \mu_4|111\rangle$ & $Z \otimes ZX \otimes ZX$ \\[2pt]
        
        $|\psi^-\rangle, |\phi^+\rangle, |\phi^+\rangle$ & 
$-\mu_4|000\rangle - \mu_5|001\rangle - \mu_6|010\rangle - \mu_7|011\rangle + \mu_0|100\rangle + \mu_1|101\rangle + \mu_2|110\rangle + \mu_3|111\rangle$ & $ZX \otimes I \otimes I$ \\
        $|\psi^-\rangle, |\phi^+\rangle, |\psi^+\rangle$ & 
$-\mu_5|000\rangle - \mu_4|001\rangle - \mu_7|010\rangle - \mu_6|011\rangle + \mu_1|100\rangle + \mu_0|101\rangle + \mu_3|110\rangle + \mu_2|111\rangle$ & $ZX \otimes I \otimes X$ \\
        $|\psi^-\rangle, |\phi^+\rangle, |\phi^-\rangle$ & 
$-\mu_4|000\rangle + \mu_5|001\rangle - \mu_6|010\rangle + \mu_7|011\rangle + \mu_0|100\rangle - \mu_1|101\rangle + \mu_2|110\rangle - \mu_3|111\rangle$ & $ZX \otimes I \otimes Z$ \\
        $|\psi^-\rangle, |\phi^+\rangle, |\psi^-\rangle$ & 
$\mu_5|000\rangle - \mu_4|001\rangle + \mu_7|010\rangle - \mu_6|011\rangle - \mu_1|100\rangle + \mu_0|101\rangle - \mu_3|110\rangle + \mu_2|111\rangle$ & $ZX \otimes I \otimes ZX$ \\[2pt]
        
        $|\psi^-\rangle, |\psi^+\rangle, |\phi^+\rangle$ & 
$-\mu_6|000\rangle - \mu_7|001\rangle - \mu_4|010\rangle - \mu_5|011\rangle + \mu_2|100\rangle + \mu_3|101\rangle + \mu_0|110\rangle + \mu_1|111\rangle$ & $ZX \otimes X \otimes I$ \\
        $|\psi^-\rangle, |\psi^+\rangle, |\psi^+\rangle$ & 
$-\mu_7|000\rangle - \mu_6|001\rangle - \mu_5|010\rangle - \mu_4|011\rangle + \mu_3|100\rangle + \mu_2|101\rangle + \mu_1|110\rangle + \mu_0|111\rangle$ & $ZX \otimes X \otimes X$ \\
        $|\psi^-\rangle, |\psi^+\rangle, |\phi^-\rangle$ & 
$-\mu_6|000\rangle + \mu_7|001\rangle - \mu_4|010\rangle + \mu_5|011\rangle + \mu_2|100\rangle - \mu_3|101\rangle + \mu_0|110\rangle - \mu_1|111\rangle$ & $ZX \otimes X \otimes Z$ \\
        $|\psi^-\rangle, |\psi^+\rangle, |\psi^-\rangle$ & 
$\mu_7|000\rangle - \mu_6|001\rangle + \mu_5|010\rangle - \mu_4|011\rangle - \mu_3|100\rangle + \mu_2|101\rangle - \mu_1|110\rangle + \mu_0|111\rangle$ & $ZX \otimes X \otimes ZX$ \\[2pt]
        
        $|\psi^-\rangle, |\phi^-\rangle, |\phi^+\rangle$ & 
$-\mu_4|000\rangle - \mu_5|001\rangle + \mu_6|010\rangle + \mu_7|011\rangle + \mu_0|100\rangle + \mu_1|101\rangle - \mu_2|110\rangle - \mu_3|111\rangle$ & $ZX \otimes Z \otimes I$ \\
        $|\psi^-\rangle, |\phi^-\rangle, |\psi^+\rangle$ & 
$-\mu_5|000\rangle - \mu_4|001\rangle + \mu_7|010\rangle + \mu_6|011\rangle + \mu_1|100\rangle + \mu_0|101\rangle - \mu_3|110\rangle - \mu_2|111\rangle$ & $ZX \otimes Z \otimes X$ \\
        $|\psi^-\rangle, |\phi^-\rangle, |\phi^-\rangle$ & 
$-\mu_4|000\rangle + \mu_5|001\rangle + \mu_6|010\rangle - \mu_7|011\rangle + \mu_0|100\rangle - \mu_1|101\rangle - \mu_2|110\rangle + \mu_3|111\rangle$ & $ZX \otimes Z \otimes Z$ \\
        $|\psi^-\rangle, |\phi^-\rangle, |\psi^-\rangle$ & 
$\mu_5|000\rangle - \mu_4|001\rangle - \mu_7|010\rangle + \mu_6|011\rangle - \mu_1|100\rangle + \mu_0|101\rangle + \mu_3|110\rangle - \mu_2|111\rangle$ & $ZX \otimes Z \otimes ZX$ \\[2pt]
        
        $|\psi^-\rangle, |\psi^-\rangle, |\phi^+\rangle$ & 
$\mu_6|000\rangle + \mu_7|001\rangle - \mu_4|010\rangle - \mu_5|011\rangle - \mu_2|100\rangle - \mu_3|101\rangle + \mu_0|110\rangle + \mu_1|111\rangle$ & $ZX \otimes ZX \otimes I$ \\
        $|\psi^-\rangle, |\psi^-\rangle, |\psi^+\rangle$ & 
$\mu_7|000\rangle + \mu_6|001\rangle - \mu_5|010\rangle - \mu_4|011\rangle - \mu_3|100\rangle - \mu_2|101\rangle + \mu_1|110\rangle + \mu_0|111\rangle$ & $ZX \otimes ZX \otimes X$ \\
        $|\psi^-\rangle, |\psi^-\rangle, |\phi^-\rangle$ & 
$\mu_6|000\rangle - \mu_7|001\rangle - \mu_4|010\rangle + \mu_5|011\rangle - \mu_2|100\rangle + \mu_3|101\rangle + \mu_0|110\rangle - \mu_1|111\rangle$ & $ZX \otimes ZX \otimes Z$ \\
        $|\psi^-\rangle, |\psi^-\rangle, |\psi^-\rangle$ & 
$-\mu_7|000\rangle + \mu_6|001\rangle + \mu_5|010\rangle - \mu_4|011\rangle + \mu_3|100\rangle - \mu_2|101\rangle - \mu_1|110\rangle + \mu_0|111\rangle$ & $ZX \otimes ZX \otimes ZX$ \\[2pt]
        
        \botrule
    \end{tabular}
\end{table}

\begin{table}[b]
    \caption{\label{tab_4}Table for each sender-receiver pair, summarizing the receiver's Bell-state measurements, along with the corresponding receiver's post-measurement states and corrective unitary operations for the \textbf{cyclic protocol} described in Sec. \ref{sec_2}.}
    \centering
    \begin{tabular}{l r l}
        \hline
        \textbf{Results} & \textbf{Post-measurement State} & \textbf{Unitaries} \\
        \hline
        $|\phi^+\rangle, |\phi^+\rangle$ & $\mu_0|00\rangle + \mu_1|01\rangle + \mu_2|10\rangle +\mu_3|11\rangle$ & $I \otimes I$ \\
        $|\phi^+\rangle, |\psi^+\rangle$ & $\mu_1|00\rangle + \mu_0|01\rangle + \mu_3|10\rangle +\mu_2|11\rangle$ & $I \otimes X$ \\
        $|\phi^+\rangle, |\phi^-\rangle$ & $\mu_0|00\rangle - \mu_1|01\rangle + \mu_2|10\rangle -\mu_3|11\rangle$ & $I \otimes Z$ \\
        $|\phi^+\rangle, |\psi^-\rangle$ & $-\mu_1|00\rangle + \mu_0|01\rangle -\mu_3|10\rangle +\mu_2|11\rangle$ & $I \otimes ZX$ \\[2pt]

        $|\psi^+\rangle, |\phi^+\rangle$ & $\mu_2|00\rangle + \mu_3|01\rangle + \mu_0|10\rangle +\mu_1|11\rangle$ & $X \otimes I$ \\
        $|\psi^+\rangle, |\psi^+\rangle$ & $\mu_3|00\rangle + \mu_2|01\rangle + \mu_1|10\rangle +\mu_0|11\rangle$ & $X \otimes X$ \\
        $|\psi^+\rangle, |\phi^-\rangle$ & $\mu_2|00\rangle - \mu_3|01\rangle + \mu_0|10\rangle -\mu_1|11\rangle$ & $X \otimes Z$ \\
        $|\psi^+\rangle, |\psi^-\rangle$ & $-\mu_3|00\rangle + \mu_2|01\rangle - \mu_1|10\rangle +\mu_0|11\rangle$ & $X \otimes ZX$ \\[2pt]

        $|\phi^-\rangle, |\phi^+\rangle$ & $\mu_0|00\rangle + \mu_1|01\rangle - \mu_2|10\rangle -\mu_3|11\rangle$ & $Z \otimes I$ \\
        $|\phi^-\rangle, |\psi^+\rangle$ & $\mu_1|00\rangle + \mu_0|01\rangle - \mu_3|10\rangle -\mu_2|11\rangle$ & $Z \otimes X$ \\
        $|\phi^-\rangle, |\phi^-\rangle$ & $\mu_0|00\rangle - \mu_1|01\rangle - \mu_2|10\rangle +\mu_3|11\rangle$ & $Z \otimes Z$ \\
        $|\phi^-\rangle, |\psi^-\rangle$ & $-\mu_1|00\rangle + \mu_0|01\rangle + \mu_3|10\rangle -\mu_2|11\rangle$ & $Z \otimes ZX$ \\[2pt]

        $|\psi^-\rangle, |\phi^+\rangle$ & $-\mu_2|00\rangle - \mu_3|01\rangle + \mu_0|10\rangle +\mu_1|11\rangle$ & $ZX \otimes I$ \\
        $|\psi^-\rangle, |\psi^+\rangle$ & $-\mu_3|00\rangle - \mu_2|01\rangle + \mu_1|10\rangle +\mu_0|11\rangle$ & $ZX \otimes X$ \\
        $|\psi^-\rangle, |\phi^-\rangle$ & $-\mu_2|00\rangle + \mu_3|01\rangle + \mu_0|10\rangle -\mu_1|11\rangle$ & $ZX \otimes Z$ \\
        $|\psi^-\rangle, |\psi^-\rangle$ & $\mu_3|00\rangle - \mu_2|01\rangle - \mu_1|10\rangle +\mu_0|11\rangle$ & $ZX \otimes ZX$ \\[2pt]
        
        \hline
    \end{tabular}
\end{table}

\section*{Appendix A. Unitary Rotation table for Bidirectional protocol}
Table \ref{tab_3} summarizes the Bell-state measurement outcomes and corresponding corrective unitaries for the bidirectional teleportation protocol. Alice performs measurements on $(a_1,1)$, $(a_2,2)$, and $(a_3,3)$, while Bob measures $(b_1,4)$, $(b_2,5)$, and $(b_3,6)$. The resulting corrections are applied by Bob on qubits $7,8,9$ and by Alice on qubits $10,11,12$, with $\mu\in{\alpha,\beta}$. Together, these results fully characterize both communication directions.

\section*{Appendix B. Unitary Rotation table for Cyclic protocol}
Table \ref{tab_4} summarizes the Bell-state measurement outcomes, collapsed states, and corresponding corrective unitaries for the cyclic teleportation protocol. Alice measures $(a_1,1)$ and $(a_2,2)$, Bob measures $(b_1,3)$ and $(b_2,4)$, and Charlie measures $(c_1,5)$ and $(c_2,6)$. The resulting corrections are applied by Bob on qubits $7,8$, Charlie on $9,10$, and Alice on $11,12$, with $\mu\in{\alpha,\beta,\gamma}$. These results completely specify the protocol.

\section*{Appendix C. Kraus operators}

\subsection*{Bit-flip Channel}
The bit-flip (BF) channel models errors that interchange the computational basis states $|0\rangle\leftrightarrow |1\rangle$ with probability $p$, while leaving the qubit unchanged with probability $(1-p)$.
The bit-flip channel is characterized by the following Kraus operators:
\begin{equation}
    K_0^{BF} = \sqrt{1 - p} \begin{bmatrix} 1 & 0 \\ 0 & 1 \end{bmatrix} = \sqrt{1 - p} ~I, \quad
    K_1^{BF} = \sqrt{p} \begin{bmatrix} 0 & 1 \\ 1 & 0 \end{bmatrix} = \sqrt{p} ~X
\end{equation}

\subsubsection*{Phase-flip Channel}
Phase-flip noise, or Pauli-$Z$ noise, is a common decoherence mechanism in quantum communication that preserves state populations while altering the relative phase between superposed components. Such phase disturbances reduce coherence and degrade teleportation fidelity, particularly for superposition states. In this model, each qubit undergoes a phase flip with probability $p$, transforming $|1\rangle \rightarrow -|1\rangle$ while leaving $|0\rangle$ unchanged. The phase-flip noise is defined by its Kraus operators as:
\begin{equation}
    K_0^{PF} = \sqrt{1 - p} \begin{bmatrix} 1 & 0 \\ 0 & 1 \end{bmatrix} = \sqrt{1 - p} ~ I, \quad
    K_1^{PF} = \sqrt{p} \begin{bmatrix} 1 & 0 \\ 0 & -1 \end{bmatrix} = \sqrt{p} ~ Z
\end{equation}

\subsubsection*{Amplitude-damping Channel}
Amplitude-damping noise is a dissipative decoherence process arising from energy relaxation, analogous to the spontaneous decay of an excited state to the ground state. Unlike phase-based errors, it directly modifies state populations by suppressing the $|1\rangle$ component while leaving $|0\rangle$ unchanged, thereby reducing coherence, entanglement, and teleportation fidelity. This relaxation process is characterized using Kraus operators as:
\begin{equation}
    K_0^{AD} = \begin{bmatrix} 1 & 0 \\ 0 & \sqrt{1 - p} \end{bmatrix}, \quad
    K_1^{AD} = \begin{bmatrix} 0 & \sqrt{p} \\ 0 & 0 \end{bmatrix}.
\end{equation}

\subsubsection*{Phase-damping Channel}
Phase-damping noise, or dephasing, arises from environmental interactions that randomize the relative phase between components of a quantum superposition without altering the populations of the computational basis states. Consequently, it suppresses the off-diagonal elements of the density matrix, leading to coherence loss while preserving state populations. In quantum teleportation, this primarily affects superposition states that depend on phase correlations, whereas basis states remain comparatively robust.
The Kraus operator for the phase-damping channel is defined as:
\begin{equation}
    K_0^{PD} = \begin{bmatrix} 1 & 0 \\ 0 & \sqrt{1 - p} \end{bmatrix}, \quad
    K_1^{PD} = \begin{bmatrix} 0 & 0 \\ 0 & \sqrt{p} \end{bmatrix}.
\end{equation}

\subsubsection*{Depolarizing Channel}
The depolarizing channel models a symmetric noise process in which environmental interactions randomly disturb the quantum state, replacing it with a maximally mixed state with probability $p$. This effect is typically represented through the equal application of Pauli errors, causing progressive loss of both classical and quantum correlations in the entangled resource and thereby reducing teleportation fidelity across all input states.
The depolarizing channel is given by the Kraus operators:
\begin{align}
    K_0^{DP} = \sqrt{1-\frac{3p}{4}} \begin{bmatrix} 1 & 0 \\ 0 & 1 \end{bmatrix} = \sqrt{1-\frac{3p}{4}} ~ I&, \quad
    K_1^{DP} = \sqrt{\frac{p}{4}} \begin{bmatrix} 0 & 1 \\ 1 & 0 \end{bmatrix} = \sqrt{\frac{p}{4}} ~ X \nonumber \\
    K_2^{DP} = \sqrt{\frac{p}{4}} \begin{bmatrix} 0 & -i \\ i & 0 \end{bmatrix} = \sqrt{\frac{p}{4}} ~ Y&, \quad
    K_3^{DP} = \sqrt{\frac{p}{4}} \begin{bmatrix} 1 & 0 \\ 0 & -1 \end{bmatrix} = \sqrt{\frac{p}{4}} ~ Z
\end{align}


\subsection*{Acknowledgements}
The authors gratefully acknowledge the support by the project ANRF/PAIR/2025/000021/PAIR-A from Anusandhan National Research Foundation (ANRF), India, and the financial support from the Council for Scientific \& Industrial Research (CSIR), India. The authors also acknowledge IBM for providing access to Qiskit, the open-source quantum computing framework that facilitated the implementation and simulation of the protocols presented in this work.

\subsection*{Declaration of competing interests}
The authors declare that they have no conflict of interest.

\subsection*{Data availability}
No data was generated for the research described in the article.

\end{appendices}

\clearpage


\begin{thebibliography}{55}
\ifx \bisbn   \undefined \def \bisbn  #1{ISBN #1}\fi
\ifx \binits  \undefined \def \binits#1{#1}\fi
\ifx \bauthor  \undefined \def \bauthor#1{#1}\fi
\ifx \batitle  \undefined \def \batitle#1{#1}\fi
\ifx \bjtitle  \undefined \def \bjtitle#1{#1}\fi
\ifx \bvolume  \undefined \def \bvolume#1{\textbf{#1}}\fi
\ifx \byear  \undefined \def \byear#1{#1}\fi
\ifx \bissue  \undefined \def \bissue#1{#1}\fi
\ifx \bfpage  \undefined \def \bfpage#1{#1}\fi
\ifx \blpage  \undefined \def \blpage #1{#1}\fi
\ifx \burl  \undefined \def \burl#1{\textsf{#1}}\fi
\ifx \doiurl  \undefined \def \doiurl#1{\url{https://doi.org/#1}}\fi
\ifx \betal  \undefined \def \betal{\textit{et al.}}\fi
\ifx \binstitute  \undefined \def \binstitute#1{#1}\fi
\ifx \binstitutionaled  \undefined \def \binstitutionaled#1{#1}\fi
\ifx \bctitle  \undefined \def \bctitle#1{#1}\fi
\ifx \beditor  \undefined \def \beditor#1{#1}\fi
\ifx \bpublisher  \undefined \def \bpublisher#1{#1}\fi
\ifx \bbtitle  \undefined \def \bbtitle#1{#1}\fi
\ifx \bedition  \undefined \def \bedition#1{#1}\fi
\ifx \bseriesno  \undefined \def \bseriesno#1{#1}\fi
\ifx \blocation  \undefined \def \blocation#1{#1}\fi
\ifx \bsertitle  \undefined \def \bsertitle#1{#1}\fi
\ifx \bsnm \undefined \def \bsnm#1{#1}\fi
\ifx \bsuffix \undefined \def \bsuffix#1{#1}\fi
\ifx \bparticle \undefined \def \bparticle#1{#1}\fi
\ifx \barticle \undefined \def \barticle#1{#1}\fi
\bibcommenthead
\ifx \bconfdate \undefined \def \bconfdate #1{#1}\fi
\ifx \botherref \undefined \def \botherref #1{#1}\fi
\ifx \url \undefined \def \url#1{\textsf{#1}}\fi
\ifx \bchapter \undefined \def \bchapter#1{#1}\fi
\ifx \bbook \undefined \def \bbook#1{#1}\fi
\ifx \bcomment \undefined \def \bcomment#1{#1}\fi
\ifx \oauthor \undefined \def \oauthor#1{#1}\fi
\ifx \citeauthoryear \undefined \def \citeauthoryear#1{#1}\fi
\ifx \endbibitem  \undefined \def \endbibitem {}\fi
\ifx \bconflocation  \undefined \def \bconflocation#1{#1}\fi
\ifx \arxivurl  \undefined \def \arxivurl#1{\textsf{#1}}\fi
\csname PreBibitemsHook\endcsname

\bibitem[\protect\citeauthoryear{Bennett et~al.}{}]{bennett_teleporting_1993}
\begin{botherref}
\oauthor{\bsnm{Bennett}, \binits{C.H.}},
\oauthor{\bsnm{Brassard}, \binits{G.}},
\oauthor{\bsnm{Crépeau}, \binits{C.}},
\oauthor{\bsnm{Jozsa}, \binits{R.}},
\oauthor{\bsnm{Peres}, \binits{A.}},
\oauthor{\bsnm{Wootters}, \binits{W.K.}}:
Teleporting an unknown quantum state via dual classical and einstein-podolsky-rosen channels
\textbf{70}(13),
1895--1899
\doiurl{10.1103/PhysRevLett.70.1895} .
\end{botherref}
\endbibitem

\bibitem[\protect\citeauthoryear{Bouwmeester et~al.}{}]{bouwmeester_experimental_1997}
\begin{botherref}
\oauthor{\bsnm{Bouwmeester}, \binits{D.}},
\oauthor{\bsnm{Pan}, \binits{J.-W.}},
\oauthor{\bsnm{Mattle}, \binits{K.}},
\oauthor{\bsnm{Eibl}, \binits{M.}},
\oauthor{\bsnm{Weinfurter}, \binits{H.}},
\oauthor{\bsnm{Zeilinger}, \binits{A.}}:
Experimental quantum teleportation
\textbf{390}(6660),
575--579.
\end{botherref}
\endbibitem

\bibitem[\protect\citeauthoryear{Nielsen and Chuang}{}]{nielsen_quantum_2010}
\begin{botherref}
\oauthor{\bsnm{Nielsen}, \binits{M.A.}},
\oauthor{\bsnm{Chuang}, \binits{I.L.}}:
Quantum Computation and Quantum Information,
10th anniversary ed edn.
Cambridge University Press
\end{botherref}
\endbibitem

\bibitem[\protect\citeauthoryear{Mishra et~al.}{}]{mishra_two-way_2011}
\begin{botherref}
\oauthor{\bsnm{Mishra}, \binits{M.K.}},
\oauthor{\bsnm{Maurya}, \binits{A.K.}},
\oauthor{\bsnm{Prakash}, \binits{H.}}:
Two-way quantum communication: ‘secure quantum information exchange’
\textbf{44}(11),
115504
\doiurl{10.1088/0953-4075/44/11/115504} .
\end{botherref}
\endbibitem

\bibitem[\protect\citeauthoryear{Zha et~al.}{}]{zha_bidirectional_2013}
\begin{botherref}
\oauthor{\bsnm{Zha}, \binits{X.-W.}},
\oauthor{\bsnm{Zou}, \binits{Z.-C.}},
\oauthor{\bsnm{Qi}, \binits{J.-X.}},
\oauthor{\bsnm{Song}, \binits{H.-Y.}}:
Bidirectional quantum controlled teleportation via five-qubit cluster state
\textbf{52}(6),
1740--1744
\doiurl{10.1007/s10773-012-1208-5} .
\end{botherref}
\endbibitem

\bibitem[\protect\citeauthoryear{Thapliyal et~al.}{}]{thapliyal_general_2015}
\begin{botherref}
\oauthor{\bsnm{Thapliyal}, \binits{K.}},
\oauthor{\bsnm{Verma}, \binits{A.}},
\oauthor{\bsnm{Pathak}, \binits{A.}}:
A general method for selecting quantum channel for bidirectional controlled state teleportation and other schemes of controlled quantum communication
\textbf{14}(12),
4601--4614
\doiurl{10.1007/s11128-015-1124-8} .
\end{botherref}
\endbibitem

\bibitem[\protect\citeauthoryear{Zhou et~al.}{}]{zhou_bidirectional_2018}
\begin{botherref}
\oauthor{\bsnm{Zhou}, \binits{S.-Q.}},
\oauthor{\bsnm{Bai}, \binits{M.-Q.}},
\oauthor{\bsnm{Liao}, \binits{T.}},
\oauthor{\bsnm{Lei}, \binits{J.}},
\oauthor{\bsnm{Tang}, \binits{L.}},
\oauthor{\bsnm{Zhang}, \binits{C.-Y.}}:
Bidirectional quantum operation teleportation with different states
\textbf{16}(5),
1850042
\doiurl{10.1142/S0219749918500429} .
\end{botherref}
\endbibitem

\bibitem[\protect\citeauthoryear{Zhou et~al.}{}]{zhou_quantum_2020}
\begin{botherref}
\oauthor{\bsnm{Zhou}, \binits{R.-G.}},
\oauthor{\bsnm{Li}, \binits{X.}},
\oauthor{\bsnm{Qian}, \binits{C.}},
\oauthor{\bsnm{Ian}, \binits{H.}}:
Quantum bidirectional teleportation 2 $\leftrightarrow$ 2 or 2 $\leftrightarrow$ 3 qubit teleportation protocol via 6-qubit entangled state
\textbf{59}(1),
166--172
\doiurl{10.1007/s10773-019-04306-1} .
\end{botherref}
\endbibitem

\bibitem[\protect\citeauthoryear{Huo et~al.}{}]{huo_controlled_2021}
\begin{botherref}
\oauthor{\bsnm{Huo}, \binits{G.}},
\oauthor{\bsnm{Zhang}, \binits{T.}},
\oauthor{\bsnm{Zha}, \binits{X.}},
\oauthor{\bsnm{Zhang}, \binits{X.}},
\oauthor{\bsnm{Zhang}, \binits{M.}}:
Controlled asymmetric bidirectional quantum teleportation of two- and three-qubit states
\textbf{20}(1),
24
\doiurl{10.1007/s11128-020-02956-3} .
\end{botherref}
\endbibitem

\bibitem[\protect\citeauthoryear{Kumar~Mandal et~al.}{}]{kumar_mandal_resumable_2024}
\begin{botherref}
\oauthor{\bsnm{Kumar~Mandal}, \binits{M.}},
\oauthor{\bsnm{Choudhury}, \binits{B.S.}},
\oauthor{\bsnm{Samanta}, \binits{S.}},
\oauthor{\bsnm{Dhara}, \binits{A.}}:
Resumable probabilistic teleportation of a three qubit state using different quantum channels
\textbf{99}(5),
055117.
\end{botherref}
\endbibitem

\bibitem[\protect\citeauthoryear{Chen et~al.}{}]{chen_cyclic_2017}
\begin{botherref}
\oauthor{\bsnm{Chen}, \binits{Y.-X.}},
\oauthor{\bsnm{Du}, \binits{J.}},
\oauthor{\bsnm{Liu}, \binits{S.-Y.}},
\oauthor{\bsnm{Wang}, \binits{X.-H.}}:
Cyclic quantum teleportation
\textbf{16}(8),
201
\doiurl{10.1007/s11128-017-1648-1} .
\end{botherref}
\endbibitem

\bibitem[\protect\citeauthoryear{Sang}{}]{sang_cyclic_2018}
\begin{botherref}
\oauthor{\bsnm{Sang}, \binits{Z.-w.}}:
Cyclic controlled teleportation by using a seven-qubit entangled state
\textbf{57}(12),
3835--3838.
\end{botherref}
\endbibitem

\bibitem[\protect\citeauthoryear{Shao and Long}{}]{shao_circular_2019}
\begin{botherref}
\oauthor{\bsnm{Shao}, \binits{Z.}},
\oauthor{\bsnm{Long}, \binits{Y.}}:
Circular controlled quantum teleportation by a genuine seven-qubit entangled state
\textbf{58}(6),
1957--1967
\doiurl{10.1007/s10773-019-04089-5} .
\end{botherref}
\endbibitem

\bibitem[\protect\citeauthoryear{Wang et~al.}{}]{wang_synchronous_2022}
\begin{botherref}
\oauthor{\bsnm{Wang}, \binits{M.-R.}},
\oauthor{\bsnm{Ren}, \binits{P.}},
\oauthor{\bsnm{Xiang}, \binits{Z.}}:
Synchronous multicast-based quantum teleportation scheme via a six-qubit entangled state
\textbf{36}(30),
2250164
\doiurl{10.1142/S0217984922501640} .
\end{botherref}
\endbibitem

\bibitem[\protect\citeauthoryear{Verma}{}]{verma_cyclic_2020}
\begin{botherref}
\oauthor{\bsnm{Verma}, \binits{V.}}:
Cyclic quantum teleportation via {GHZ}-like state
\textbf{35}(40),
2050333
\doiurl{10.1142/S0217732320503332} .
\end{botherref}
\endbibitem

\bibitem[\protect\citeauthoryear{Verma}{}]{verma_cyclic_2021}
\begin{botherref}
\oauthor{\bsnm{Verma}, \binits{V.}}:
Cyclic quantum teleportation via \textit{G} -states
\textbf{35}(8),
2150145
\doiurl{10.1142/S0217984921501451} .
\end{botherref}
\endbibitem

\bibitem[\protect\citeauthoryear{Slaoui et~al.}{}]{slaoui_cyclic_2024}
\begin{botherref}
\oauthor{\bsnm{Slaoui}, \binits{A.}},
\oauthor{\bsnm{El~Kirdi}, \binits{M.}},
\oauthor{\bsnm{Ahl~Laamara}, \binits{R.}},
\oauthor{\bsnm{Alabdulhafith}, \binits{M.}},
\oauthor{\bsnm{Chelloug}, \binits{S.A.}},
\oauthor{\bsnm{Abd~El-Latif}, \binits{A.A.}}:
Cyclic quantum teleportation of two-qubit entangled states by using six-qubit cluster state and six-qubit entangled state
\textbf{14}(1),
15856.
\end{botherref}
\endbibitem

\bibitem[\protect\citeauthoryear{Yang}{}]{yang_quantum_2022}
\begin{botherref}
\oauthor{\bsnm{Yang}, \binits{B.}}:
Quantum asymmetric cyclic teleportation of arbitrary single-particle and two-particle states
\textbf{36}(26),
2250142
\doiurl{10.1142/S0217984922501421} .
\end{botherref}
\endbibitem

\bibitem[\protect\citeauthoryear{Yang}{}]{yang_asymmetric_2022}
\begin{botherref}
\oauthor{\bsnm{Yang}, \binits{B.}}:
Asymmetric cyclic controlled quantum teleportation by using a quantum channel composed of six g states
\textbf{36}(22),
2250113
\doiurl{10.1142/S0217984922501135} .
\end{botherref}
\endbibitem

\bibitem[\protect\citeauthoryear{Choudhury and Samanta}{}]{choudhury_controlled_2020}
\begin{botherref}
\oauthor{\bsnm{Choudhury}, \binits{B.S.}},
\oauthor{\bsnm{Samanta}, \binits{S.}}:
A controlled protocol for asymmetric cyclic (a $\Rightarrow$ b $\Rightarrow$ c $\Rightarrow$ a) quantum state transfer between three parties
\textbf{95}(1),
015101.
\end{botherref}
\endbibitem

\bibitem[\protect\citeauthoryear{Verma}{}]{verma_symmetric_2021}
\begin{botherref}
\oauthor{\bsnm{Verma}, \binits{V.}}:
Symmetric and asymmetric cyclic controlled quantum teleportation via nine-qubit entangled state
\textbf{35}(15),
2150249
\doiurl{10.1142/S0217984921502493} .
\end{botherref}
\endbibitem

\bibitem[\protect\citeauthoryear{Kaur et~al.}{}]{kaur_asymmetric_2024}
\begin{botherref}
\oauthor{\bsnm{Kaur}, \binits{S.}},
\oauthor{\bsnm{Lal}, \binits{J.}},
\oauthor{\bsnm{Gill}, \binits{S.}}:
Asymmetric controlled cyclic quantum teleportation of two, three and four qubit states with optimal quantum resources
\textbf{23}(4),
141
\doiurl{10.1007/s11128-024-04345-6} .
\end{botherref}
\endbibitem

\bibitem[\protect\citeauthoryear{Zha and Miao}{}]{zha_hierarchical_2019}
\begin{botherref}
\oauthor{\bsnm{Zha}, \binits{X.-W.}},
\oauthor{\bsnm{Miao}, \binits{N.}}:
Hierarchical controlled quantum teleportation
\textbf{33}(29),
1950356
\doiurl{10.1142/S0217984919503561} .
\end{botherref}
\endbibitem

\bibitem[\protect\citeauthoryear{Zha}{}]{zha_different_2025}
\begin{botherref}
\oauthor{\bsnm{Zha}, \binits{X.-W.}}:
Different level of controlled cyclic quantum teleportation
\textbf{99}(7),
2581--2585
\doiurl{10.1007/s12648-024-03475-y} .
\end{botherref}
\endbibitem

\bibitem[\protect\citeauthoryear{Zha and Li}{}]{zha_four-directional_2020}
\begin{botherref}
\oauthor{\bsnm{Zha}, \binits{X.-W.}},
\oauthor{\bsnm{Li}, \binits{K.}}:
Four-directional quantum controlled teleportation using a single quantum resource
\textbf{34}(35),
2050412
\doiurl{10.1142/S0217984920504126} .
\end{botherref}
\endbibitem

\bibitem[\protect\citeauthoryear{Fatahi and Parsamehr}{}]{fatahi_multi-hop_2024}
\begin{botherref}
\oauthor{\bsnm{Fatahi}, \binits{N.}},
\oauthor{\bsnm{Parsamehr}, \binits{S.}}:
Multi-hop quantum teleportation of a \textit{N} -qubit quantum state by using cluster state
\textbf{38}(23),
2450191
\doiurl{10.1142/S0217984924501914} .
\end{botherref}
\endbibitem

\bibitem[\protect\citeauthoryear{Saha et~al.}{}]{saha_hybrid_2025}
\begin{botherref}
\oauthor{\bsnm{Saha}, \binits{P.}},
\oauthor{\bsnm{Choudhury}, \binits{B.S.}},
\oauthor{\bsnm{Metwally}, \binits{N.}},
\oauthor{\bsnm{Mandal}, \binits{M.K.}}:
Hybrid of quantum teleportation and remote state preparation protocol for different choices of receivers with hierarchy
\textbf{563},
131041
\doiurl{10.1016/j.physleta.2025.131041} .
\end{botherref}
\endbibitem

\bibitem[\protect\citeauthoryear{Paulson and Satyanarayana}{}]{paulson_bounds_2017}
\begin{botherref}
\oauthor{\bsnm{Paulson}, \binits{K.G.}},
\oauthor{\bsnm{Satyanarayana}, \binits{S.V.M.}}:
Bounds on mixedness and entanglement of quantum teleportation resources
\textbf{381}(13),
1134--1137
\doiurl{10.1016/j.physleta.2017.02.010} .
\end{botherref}
\endbibitem

\bibitem[\protect\citeauthoryear{Zhou et~al.}{}]{zhou_cyclic_2019}
\begin{botherref}
\oauthor{\bsnm{Zhou}, \binits{R.-G.}},
\oauthor{\bsnm{Qian}, \binits{C.}},
\oauthor{\bsnm{Ian}, \binits{H.}}:
Cyclic and bidirectional quantum teleportation via pseudo multi-qubit states
\textbf{7},
42445--42449.
\end{botherref}
\endbibitem

\bibitem[\protect\citeauthoryear{Wu et~al.}{}]{wu_cyclic_2020}
\begin{botherref}
\oauthor{\bsnm{Wu}, \binits{F.}},
\oauthor{\bsnm{Bai}, \binits{M.-Q.}},
\oauthor{\bsnm{Zhang}, \binits{Y.-C.}},
\oauthor{\bsnm{Liu}, \binits{R.-J.}},
\oauthor{\bsnm{Mo}, \binits{Z.-W.}}:
Cyclic quantum teleportation of an unknown multi-particle high-dimension state
\textbf{34}(5),
2050073
\doiurl{10.1142/S0217984920500736} .
\end{botherref}
\endbibitem

\bibitem[\protect\citeauthoryear{Verma}{}]{verma_bidirectional_2020}
\begin{botherref}
\oauthor{\bsnm{Verma}, \binits{V.}}:
Bidirectional quantum teleportation and cyclic quantum teleportation of multi-qubit entangled states via g-state
\textbf{34}(28),
2050261
\doiurl{10.1142/S0217979220502616} .
\end{botherref}
\endbibitem

\bibitem[\protect\citeauthoryear{Xu and Zhou}{}]{xu_asymmetric_2023}
\begin{botherref}
\oauthor{\bsnm{Xu}, \binits{J.}},
\oauthor{\bsnm{Zhou}, \binits{R.-G.}}:
Asymmetric bidirectional cyclic controlled quantum teleportation in noisy environment
\textbf{22}(10),
376
\doiurl{10.1007/s11128-023-04116-9} .
\end{botherref}
\endbibitem

\bibitem[\protect\citeauthoryear{Choudhury et~al.}{}]{choudhury_bi-directional_2023}
\begin{botherref}
\oauthor{\bsnm{Choudhury}, \binits{B.S.}},
\oauthor{\bsnm{Mandal}, \binits{M.K.}},
\oauthor{\bsnm{Dolai}, \binits{B.}},
\oauthor{\bsnm{Samanta}, \binits{S.}}:
Bi-directional controlled asymmetric teleportation protocol initiated by a mentor in noisy environments
\textbf{98}(9),
095107.
\end{botherref}
\endbibitem

\bibitem[\protect\citeauthoryear{Liu et~al.}{}]{liu_cyclic-controlled_2019}
\begin{botherref}
\oauthor{\bsnm{Liu}, \binits{R.-J.}},
\oauthor{\bsnm{Bai}, \binits{M.-Q.}},
\oauthor{\bsnm{Wu}, \binits{F.}},
\oauthor{\bsnm{Zhang}, \binits{Y.-C.}}:
Cyclic-controlled quantum operation teleportation in noisy environment
\textbf{17}(7),
1950052
\doiurl{10.1142/S0219749919500527} .
\end{botherref}
\endbibitem

\bibitem[\protect\citeauthoryear{Kaur et~al.}{}]{kaur_multidirectional_2023}
\begin{botherref}
\oauthor{\bsnm{Kaur}, \binits{S.}},
\oauthor{\bsnm{{Priyanka}}},
\oauthor{\bsnm{Lal}, \binits{J.}},
\oauthor{\bsnm{Gill}, \binits{S.}}:
Multidirectional quantum controlled teleportation in noisy environment
\textbf{62}(11),
249
\doiurl{10.1007/s10773-023-05472-z} .
\end{botherref}
\endbibitem

\bibitem[\protect\citeauthoryear{Sisodia et~al.}{}]{sisodia_hybrid_2024}
\begin{botherref}
\oauthor{\bsnm{Sisodia}, \binits{M.}},
\oauthor{\bsnm{Mandal}, \binits{M.K.}},
\oauthor{\bsnm{Choudhury}, \binits{B.S.}}:
Hybrid multi-directional quantum communication protocol
\textbf{23}(9),
310
\doiurl{10.1007/s11128-024-04516-5} .
\end{botherref}
\endbibitem

\bibitem[\protect\citeauthoryear{Yang et~al.}{}]{yang_bidirectional_2017}
\begin{botherref}
\oauthor{\bsnm{Yang}, \binits{G.}},
\oauthor{\bsnm{Lian}, \binits{B.-W.}},
\oauthor{\bsnm{Nie}, \binits{M.}},
\oauthor{\bsnm{Jin}, \binits{J.}}:
Bidirectional multi-qubit quantum teleportation in noisy channel aided with weak measurement
\textbf{26}(4),
040305.
\end{botherref}
\endbibitem

\bibitem[\protect\citeauthoryear{Mandal et~al.}{}]{mandal_quantum_2024}
\begin{botherref}
\oauthor{\bsnm{Mandal}, \binits{M.K.}},
\oauthor{\bsnm{Choudhury}, \binits{B.S.}},
\oauthor{\bsnm{Samanta}, \binits{S.}}:
Quantum teleportation of w-type states in the presence of a controller
\textbf{38}(3),
2350232
\doiurl{10.1142/S0217984923502329} .
\end{botherref}
\endbibitem

\bibitem[\protect\citeauthoryear{Kaur and Gill}{}]{kaur_efficient_2024}
\begin{botherref}
\oauthor{\bsnm{Kaur}, \binits{S.}},
\oauthor{\bsnm{Gill}, \binits{S.}}:
Efficient high-dimensional cyclic quantum teleportation via a 3-dimensional {GHZ} state in a noisy environment
\textbf{90},
956--969
\doiurl{10.1016/j.cjph.2024.06.022} .
\end{botherref}
\endbibitem

\bibitem[\protect\citeauthoryear{Taufiqi et~al.}{}]{taufiqi_teleportation_2025}
\begin{botherref}
\oauthor{\bsnm{Taufiqi}, \binits{M.}},
\oauthor{\bsnm{Yuwana}, \binits{L.}},
\oauthor{\bsnm{Muniandy}, \binits{S.V.}},
\oauthor{\bsnm{Artawan}, \binits{I.N.}},
\oauthor{\bsnm{Rahmawati}, \binits{R.}},
\oauthor{\bsnm{Subagyo}, \binits{B.A.}},
\oauthor{\bsnm{Sukamto}, \binits{H.}},
\oauthor{\bsnm{Purwanto}, \binits{A.}}:
On the teleportation superiority in noisy environments
\textbf{64}(10),
262
\doiurl{10.1007/s10773-025-06108-0} .
\end{botherref}
\endbibitem

\bibitem[\protect\citeauthoryear{Yuan et~al.}{}]{yuan_optimizing_2008}
\begin{botherref}
\oauthor{\bsnm{Yuan}, \binits{H.}},
\oauthor{\bsnm{Liu}, \binits{Y.-M.}},
\oauthor{\bsnm{Zhang}, \binits{W.}},
\oauthor{\bsnm{Zhang}, \binits{Z.-J.}}:
Optimizing resource consumption, operation complexity and efficiency in quantum-state sharing
\textbf{41}(14),
145506
\doiurl{10.1088/0953-4075/41/14/145506} .
\end{botherref}
\endbibitem

\bibitem[\protect\citeauthoryear{Li and Jin}{}]{li_bidirectional_2016}
\begin{botherref}
\oauthor{\bsnm{Li}, \binits{Y.-h.}},
\oauthor{\bsnm{Jin}, \binits{X.-m.}}:
Bidirectional controlled teleportation by using nine-qubit entangled state in noisy environments
\textbf{15}(2),
929--945
\doiurl{10.1007/s11128-015-1194-7} .
\end{botherref}
\endbibitem

\bibitem[\protect\citeauthoryear{Zhou et~al.}{}]{zhou_asymmetric_2019}
\begin{botherref}
\oauthor{\bsnm{Zhou}, \binits{R.-G.}},
\oauthor{\bsnm{Zhang}, \binits{Y.-N.}},
\oauthor{\bsnm{Xu}, \binits{R.}},
\oauthor{\bsnm{Qian}, \binits{C.}},
\oauthor{\bsnm{Hou}, \binits{I.}}:
Asymmetric bidirectional controlled teleportation by using nine-qubit entangled state in noisy environment
\textbf{7},
75247--75264.
\end{botherref}
\endbibitem

\bibitem[\protect\citeauthoryear{Jiang et~al.}{}]{jiang_bidirectional_2021}
\begin{botherref}
\oauthor{\bsnm{Jiang}, \binits{Y.-L.}},
\oauthor{\bsnm{Zhou}, \binits{R.-G.}},
\oauthor{\bsnm{Hao}, \binits{D.-Y.}},
\oauthor{\bsnm{Hu}, \binits{W.}}:
Bidirectional controlled quantum teleportation of three-qubit state by a new entangled eleven-qubit state
\textbf{60}(9),
3618--3630
\doiurl{10.1007/s10773-021-04935-5} .
\end{botherref}
\endbibitem

\bibitem[\protect\citeauthoryear{Dai and Li}{}]{dai_asymmetric_2022}
\begin{botherref}
\oauthor{\bsnm{Dai}, \binits{R.}},
\oauthor{\bsnm{Li}, \binits{H.-S.}}:
Asymmetric bidirectional quantum teleportation via seven-qubit cluster state
\textbf{61}(7),
187
\doiurl{10.1007/s10773-022-05157-z} .
\end{botherref}
\endbibitem

\bibitem[\protect\citeauthoryear{Wang et~al.}{}]{wang_bidirectional_2022}
\begin{botherref}
\oauthor{\bsnm{Wang}, \binits{M.-R.}},
\oauthor{\bsnm{Xiang}, \binits{Z.}},
\oauthor{\bsnm{Ren}, \binits{P.}}:
Bidirectional controlled quantum teleportation of arbitrary two-qubit states using ten-qubit entangled channel in noisy environment
\textbf{61}(11),
259
\doiurl{10.1007/s10773-022-05229-0} .
\end{botherref}
\endbibitem

\bibitem[\protect\citeauthoryear{Choudhury et~al.}{}]{choudhury_bidirectional_2023}
\begin{botherref}
\oauthor{\bsnm{Choudhury}, \binits{B.S.}},
\oauthor{\bsnm{Mandal}, \binits{M.K.}},
\oauthor{\bsnm{Samanta}, \binits{S.}},
\oauthor{\bsnm{Dolai}, \binits{B.}}:
A bidirectional hybrid quantum communication scheme for a known and an unknown qubit
\textbf{10}(1),
89--99
\doiurl{10.1007/s40509-022-00284-y} .
\end{botherref}
\endbibitem

\bibitem[\protect\citeauthoryear{Li et~al.}{}]{li_tripartite_2016}
\begin{botherref}
\oauthor{\bsnm{Li}, \binits{W.}},
\oauthor{\bsnm{Zha}, \binits{X.-W.}},
\oauthor{\bsnm{Qi}, \binits{J.-X.}}:
Tripartite quantum controlled teleportation via seven-qubit cluster state
\textbf{55}(9),
3927--3933
\doiurl{10.1007/s10773-016-3022-y} .
\end{botherref}
\endbibitem

\bibitem[\protect\citeauthoryear{Verma et~al.}{}]{verma_improvement_2021}
\begin{botherref}
\oauthor{\bsnm{Verma}, \binits{V.}},
\oauthor{\bsnm{Yadav}, \binits{D.}},
\oauthor{\bsnm{Mishra}, \binits{D.K.}}:
Improvement on cyclic controlled teleportation by using a seven-qubit entangled state
\textbf{53}(8),
448
\doiurl{10.1007/s11082-021-03098-1} .
\end{botherref}
\endbibitem

\bibitem[\protect\citeauthoryear{Zhou and Ling}{}]{zhou_asymmetric_2021}
\begin{botherref}
\oauthor{\bsnm{Zhou}, \binits{R.-G.}},
\oauthor{\bsnm{Ling}, \binits{C.}}:
Asymmetric cyclic controlled quantum teleportation by using nine-qubit entangled state
\textbf{60}(9),
3435--3459
\doiurl{10.1007/s10773-021-04825-w} .
\end{botherref}
\endbibitem

\bibitem[\protect\citeauthoryear{Rahmawati et~al.}{}]{rahmawati_symmetric_2022}
\begin{botherref}
\oauthor{\bsnm{Rahmawati}, \binits{R.}},
\oauthor{\bsnm{Purwanto}, \binits{A.}},
\oauthor{\bsnm{Subagyo}, \binits{B.A.}},
\oauthor{\bsnm{Taufiqi}, \binits{M.}},
\oauthor{\bsnm{Hatmoko}, \binits{B.D.}}:
Symmetric and asymmetric cyclic quantum teleportation with different controller for each participant
\textbf{61}(10),
244
\doiurl{10.1007/s10773-022-05208-5} .
\end{botherref}
\endbibitem

\bibitem[\protect\citeauthoryear{Mahjoory et~al.}{}]{mahjoory_asymmetric_2023}
\begin{botherref}
\oauthor{\bsnm{Mahjoory}, \binits{A.}},
\oauthor{\bsnm{Kazemikhah}, \binits{P.}},
\oauthor{\bsnm{Aghababa}, \binits{H.}},
\oauthor{\bsnm{Kolahdouz}, \binits{M.}}:
Asymmetric tridirectional quantum teleportation using seven-qubit cluster states
\textbf{98}(8),
085218.
\end{botherref}
\endbibitem

\bibitem[\protect\citeauthoryear{Jiang and Shi}{}]{jiang_multi-party_2024}
\begin{botherref}
\oauthor{\bsnm{Jiang}, \binits{S.-X.}},
\oauthor{\bsnm{Shi}, \binits{J.}}:
Multi-party three-dimensional asymmetric cyclic controlled quantum teleportation in noisy environment
\textbf{23}(7),
261
\doiurl{10.1007/s11128-024-04474-y} .
\end{botherref}
\endbibitem

\bibitem[\protect\citeauthoryear{Taufiqi et~al.}{}]{taufiqi_cyclic_2024}
\begin{botherref}
\oauthor{\bsnm{Taufiqi}, \binits{M.}},
\oauthor{\bsnm{Purwanto}, \binits{A.}},
\oauthor{\bsnm{Subagyo}, \binits{B.A.}},
\oauthor{\bsnm{Hatmoko}, \binits{B.D.}}:
Cyclic quantum teleportation with multi-level of control
\textbf{63}(1),
9
\doiurl{10.1007/s10773-023-05513-7} .
\end{botherref}
\endbibitem

\bibitem[\protect\citeauthoryear{Kaur and Gill}{}]{kaur_asymmetric_2025}
\begin{botherref}
\oauthor{\bsnm{Kaur}, \binits{S.}},
\oauthor{\bsnm{Gill}, \binits{S.}}:
Asymmetric quad-directional controlled quantum teleportation in noisy environment
\textbf{25}(1),
1
\doiurl{10.1007/s11128-025-04981-6} .
\end{botherref}
\endbibitem

\end{thebibliography}


\end{document}